\documentclass[12pt,a4paper]{article}
\pdfoutput=1
\usepackage{jheppub}

\usepackage{graphicx, epsfig}
\usepackage{color}
\usepackage{mathrsfs}
\usepackage{amsmath}
\usepackage{amssymb}

\newcommand{\order}{\mathcal{O}}
\newcommand{\tphi}{\tilde \phi}
\newcommand{\tp}{\tilde p}
\newcommand{\del}{\partial}
\newcommand{\Vcan}{V_{can}}

\newcommand{\be}{\begin{equation}}
\newcommand{\ee}{\end{equation}}
\newcommand{\bea}{\begin{eqnarray}}
\newcommand{\eea}{\end{eqnarray}}
\newcommand{\barr}{\begin{array}}
\newcommand{\earr}{\end{array}}
\newcommand {\lessim} {\ {\raise-.5ex\hbox{$\buildrel<\over\sim$}}\ }

\def\beq{\begin{equation}}
\def\eeq{\end{equation}}
\def\be{\begin{equation}}
\def\ee{\end{equation}}
\def\bea{\begin{eqnarray}}
\def\eea{\end{eqnarray}}

\title{Relations between canonical and non-canonical inflation}
\author[a]{Rhiannon~Gwyn,} \author[b]{Markus~Rummel} \author[c]{and Alexander~Westphal}

\affiliation[a]{AEI Max-Planck-Institut f\"ur Gravitationsphysik, D-14476 Potsdam, Germany}
\affiliation[b]{II. Institut f\"ur Theoretische Physik der Universit\"at Hamburg, D-22761 Hamburg, Germany}
\affiliation[c]{Deutsches Elektronen-Synchrotron DESY, Theory Group, D-22603 Hamburg, Germany}
\emailAdd{rhiannon.gwyn@aei.mpg.de}\emailAdd{markus.rummel@desy.de}\emailAdd{alexander.westphal@desy.de}
\abstract{We look for potential observational degeneracies between canonical and non-canonical models of inflation of a single field $\phi$. Non-canonical inflationary models are characterized by higher than linear powers of the standard kinetic term $X$ in the effective Lagrangian $p(X,\phi)$ and arise for instance in the context of the Dirac-Born-Infeld (DBI) action in string theory. An on-shell transformation is introduced that transforms non-canonical inflationary theories to theories with a canonical kinetic term. The 2-point function observables of the original non-canonical theory and its canonical transform are found to match in the case of DBI inflation.}

\begin{document}
\noindent December 17, 2012\hfill
DESY 12-245
\\[0.7cm]
\maketitle
\setlength{\parskip}{0.2cm} 


\section{Introduction}
Since the first conclusive detection~\cite{Smoot:1992td} of the $\Delta T/T={\cal O}(10^{-5})$ temperature fluctuations in the cosmic microwave background radiation (CMB), a new concordance picture of cosmology has been established. This is supported by  vastly increased observational precision in CMB measurements~\cite{Komatsu:2010fb,Dunkley:2010ge,Story:2012wx,Ade:2013uln,Ade:2013zuv} as well as measurements of the redshift-distance relation for large samples of distant type IA supernovae~\cite{Riess:1998cb,Perlmutter:1998np}, baryon acoustic oscillations (BAO)~\cite{Percival:2009xn}, and the Hubble parameter $H_0$ by the Hubble Space Telescope key project~\cite{Riess:2011yx}. The results paint a Universe very close to being spatially flat, where large-scale structure originates from a pattern of coherent acoustic oscillations in the early dense plasma which was seeded by an almost scale-invariant power spectrum of super-horizon size curvature perturbations with Gaussian distribution. These initial 
conditions arise as a direct 
consequence of a wide class of models of cosmological inflation driven by the potential energy of a scalar field. An inflationary origin of the observed curvature perturbation spectrum predicts in addition the existence of a similarly almost scale-invariant power spectrum of super-horizon size primordial gravitational waves. The magnitude of this `tensor mode' power spectrum, and in turn its detectability, is determined by the energy scale at which inflation took place.

Such almost scale-invariant power spectra of super-horizon size curvature perturbations and tensor mode perturbations with Gaussian distribution can be described at the Gaussian level by just three observational quantities: The overall normalization $\Delta_s^2$ of the curvature perturbation power spectrum (known since the COBE measurements~\cite{Smoot:1992td}), its spectral tilt $n_s$, describing the (small) deviations from scale invariance expected in most models of inflation, and the fractional power $r$ in tensor modes. $n_s$ has been constrained by various combinations of the WMAP satellite CMB results~\cite{Komatsu:2010fb} with type IA SN, BAO and $H_0$ data. In combination with the recently released ca. two-and-a-half full-sky surveys of PLANCK CMB temperature data~\cite{Ade:2013uln,Ade:2013zuv}, and earlier the 2012 Atacama Cosmology Telescope~\cite{Dunkley:2010ge}, and South Pole Telescope CMB data~\cite{Story:2012wx}, this led to an unambiguous  $>5$-$\sigma$ detection of a red tilt $n_s<1$. The 
tensor mode power fraction $r$ is so far subject to an upper bound, most recently improved to $r<0.12\,(95\%)$ by the PLANCK analysis~\cite{Ade:2013uln,Ade:2013zuv}. A future analysis of 
data of the PLANCK satellite CMB B-mode polarization results as well as future polarized ground-based CMB detectors may substantially sharpen this upper bound in the next few years.

Inflationary theory determines these three numbers in terms of the value of the scalar potential $V_0$ at the time when the largest observable scales exited the inflationary horizon (about 60 e-folds before the end of inflation), and its first and second derivatives $V'_0$, $V''_0$ with respect to the inflation scalar field $\phi$ at that time. This implies that there are huge classes of scalar potentials $V(\phi)$ even for single-field models which yield identical predictions for $\Delta_s^2$, $n_s$, and $r$.

In any attempt to connect data with theory, potential degeneracies must be taken into account before any conclusions can be drawn. In this context it is important to understand the structure of this very large model space, and look for degeneracies between large classes of inflationary models with respect to the three observable quantities. We will restrict our attention here to single-field models of inflation which partition into two large classes: models with a canonically normalized kinetic term $\frac12 (\partial_\mu\phi)^2$, and so-called non-canonical inflation models with Lagrangian ${\cal L}=p\left((\partial_\mu\phi)^2,\phi\right)$. Non-canonical inflation has been studied field-theoretically in the context of k-inflation~\cite{Garriga:1999vw}, and within string theory in DBI-inflation~\cite{Silverstein:2003hf}. In both cases the function $p\left((\partial_\mu\phi)^2,\phi\right)$ can be written as an (infinite) sum over higher powers of the derivative $(\partial_\mu\phi)^2$ with potentially field-
dependent pre-factors. These terms can lead to additional effective friction terms in the equations of motion for the inflaton. They 
can slow down the rolling of the scalar field into a regime of vacuum energy domination for potentials which would be too steep to do so in presence of a canonically normalized kinetic term alone. More general studies of such non-canonical models of inflation can be found in~\cite{Franche:2009gk}, while the effective field theory of inflationary quantum fluctuations in such general settings is discussed in~\cite{Cheung:2007st}. Non-canonical inflation quite generally leads to appreciable levels of non-Gaussianity of the inflationary quantum fluctuations~\cite{Garriga:1999vw,Silverstein:2003hf}, which has been analyzed more generally in~\cite{Chen:2006nt}, and has its full effective field theory treatment in~\cite{Cheung:2007st}.

We will look at the question of whether there are degeneracies between canonical and non-canonical models of inflation with respect to the three observational quantities describing their predicted power spectra at the Gaussian level. This question has been attacked from the point of view of reconstructing the inflationary action from observables using Monte Carlo simulations in \cite{Easson:2012id}.\footnote{Earlier work towards reconstructing the inflationary potential was done for a canonical scalar field in \cite{Easther:2002rw}, and for  a general action with noncanonical kinetic terms in \cite{Bean:2008ga, Gauthier:2008mq, Powell:2008bi}.}

The method of canonical transformations for transforming noncanonical kinetic terms into canonical kinetic terms, even in $0+1$D, appears to be limited to the case where the noncanonical theory has a quadratic potential, as we elucidate in Appendix \ref{sec:AppB}. Therefore we work here at the level of the action and of the inflationary solution itself. While formally non-canonical 2-derivative models of the form ${\cal L}=f(\phi)(\partial_\mu\phi)^2-V(\phi)$ can always be transformed off-shell by a local field redefinition into a canonical model with a transformed scalar potential, this question is rather non-trivial in the presence of 
higher-power kinetic terms. As the inflationary behavior of a given model is described in terms of a generalized slow-roll attractor solution in phase space, we will look at possible on-shell transformations of a given non-canonical model on its inflationary attractor into an equivalent canonical slow-roll inflation model. We find the general formalism for performing this matching of trajectories, which will give the canonical potential $V_{can}$ leading to slow-roll inflation in a canonical theory, with inflationary trajectory $X_{inf} (\phi)$ matching exactly that in the given non-canonical model. This matching is quite general. 

Furthermore, the 2-point observables $\Delta_s^2$, $n_s$, and $r$ are shown, numerically and analytically, to match in the case of DBI inflation, over a range of efolds. This degeneracy is nontrivial, and seen for a large range of field values well outside of the canonical regime of DBI. It could not be resolved with the currently available data at the 2-point level, requiring a measurement of the ratio of $r$ and $n_T$ to distinguish the two theories.
Note that 3-point function observables, i.e. non-Gaussianities, while generally negligible in single-field canonical inflationary models, can be appreciable in certain special cases. A sum of oscillating terms in the potential can lead to an approximately equilateral-type non-Gaussian signal \cite{Gwyn:2012pb}, while coupling of the inflaton to gauge quanta can also give rise to large equilateral-type non-Gaussianity \cite{Barnaby:2010vf, Barnaby:2011qe}. \footnote{Note that such models may be subject to a strong bound on the power spectrum coming from the non-detection of primordial black holes \cite{Linde:2012bt}.} This becomes even more interesting given that the analysis of the recent PLANCK CMB temperature data constrained local-shape non-Gaussianity arising from multi-field inflation models with $f_{NL}^{loc.}=2.7\pm 5.8$~\cite{Ade:2013ydc} down to non-primordial foreground levels, while leaving a considerable window for equilateral non-Gaussianity with $f_{NL}^{equil.}=42\pm75$~\cite{Ade:2013ydc}.
Hence, a matching of the 2-point function observables can in principle be extended to 3-point function observables by adding additional couplings or features to the potential of the canonical theory.  We find matching of the 2-point function observables to be possible precisely for the case of DBI inflation while failing for simple classes of DBI-inspired generalizations. This may point to a special status for DBI inflation as a member of the non-canonical class in that it can be related to a canonical model of inflation with matching 2-point function observables.

Our discussion proceeds as follows. In Section~\ref{sec:review} we review briefly the relevant aspects of non-canonical inflation, while in Section~\ref{findcanpot_sec} we discuss the on-shell transformation of a non-canonical model into a canonical one on the inflationary attractor of the non-canonical model. Section~\ref{theoriesspeedlimit_sec} discusses the relation of the 2-point function observables under the transformation between several classes of non-canonical inflation with a speed limit inspired by and including DBI inflation and their associated canonical models. Our main example, DBI inflation, we analyze in Section~\ref{DBIinflectionexample_sec}. Section~\ref{csnot1_sec} treats the corrections from typically the reduced speed of sound in non-canonical inflation to the 2-point function observables, and we conclude in Section~\ref{sec:conc}. There are two appendices which contain a short discussion of the accessibility of the non-canonical regime for DBI inflation (Appendix~\ref{sec:AppA}), and 
an analysis of possible off-shell transformations between non-canonical and canonical theories using a form of canonical transformations (Appendix~\ref{sec:AppB}).

\section{Review: Non-canonical inflation}\label{sec:review}

We study inflationary dynamics of a single scalar field $\phi$ minimally coupled to gravity via
\begin{equation}
 S = \int d^4 x \sqrt{g} \left[\frac{M_p^2}{2}\mathcal{R} + p(X,\phi) \right]\,,\label{Leffgen}
\end{equation}
with $X\equiv -(\partial_\mu \phi)^2 =  \dot \phi^2/2$ in a homogeneous FLRW background $ds^2 = -dt^2 + a(t)^2 dx^2$.
 
From an effective field theory point of view, we expect the function $p(X,\phi)$ to have the form
\begin{equation}
 p(X,\phi)=\sum_{n\geq 0} c_n(\phi) \frac{X^{n+1}}{\Lambda^{4n}} - V(\phi) = \Lambda^4 S(X,\phi)-V(\phi)\,,\label{psum}
\end{equation}
with some cutoff scale $\Lambda$. In this work, we will restrict ourselves to the case where the coefficients $c_n$ are not field dependent, i.e. $c_n(\phi)=c_n$, such that $p(X,\phi)$ is separable, i.e.
\begin{equation}
 p(X,\phi)= \Lambda^4 S(X) - V(\phi)\,.
\end{equation}
A theory is intrinsically non-canonical if the higher order kinetic terms $X^n$ with $n>1$ play a significant role in the dynamics. Note that this qualitatively different from theories with non-canonical kinetic terms where one can at least in principle find a field redefinition transforming to a canonical theory.

The inflationary dynamics and observables are described in terms of the generalized slow-roll parameters~\cite{Garriga:1999vw,Chen:2006nt} given as
\begin{equation}
 H=\frac{\dot{a}}{a}\,,\quad \epsilon=-\frac{\dot{H}}{H^2}\,,\quad \eta=\frac{\dot \epsilon}{H\,\epsilon}\,,\quad \kappa=\frac{\dot c_s}{H\,c_s}\,,\quad c_s^2 = \left(1+2X\frac{\partial^2 p / \partial X^2}{\partial p / \partial X} \right)^{-1}\,,
\end{equation}
which reduce to
\begin{equation}
 H=\frac{\dot{a}}{a}\,,\quad \epsilon=\epsilon_{V}=\frac{1}{2} \left(\frac{V'}{V}\right)^2\,,\quad \eta= 4 \epsilon_{V} - 2\eta_{V}\,, \quad \eta_{V} = \frac{V''}{V}\,,\quad \kappa=0\,,\quad c_s^2 = 1\,.
\end{equation}
in the canonical case $p(X,\phi)= X - V(\phi)$. The equations of motion can be derived as the Friedmann equations of a perfect fluid
\begin{align}
\begin{aligned}
&H^2 = \frac{1}{3M_p^2}\rho\,,\\
&\frac{\ddot a}{a} = - \frac{1}{6 M_p^2} (\rho + 3p)\,,\label{Friedmanneq}
\end{aligned}
\end{align}
with pressure $p = p(X,\phi)$ and energy density
\begin{equation}
 \rho = 2 X \frac{\partial p}{\partial X} - p\,.
\end{equation}
Inflationary solutions $p_{inf}\simeq - \rho_{inf}$ to eq.~\eqref{Friedmanneq} can be found as algebraic solutions $X_{inf}=X(A)$ to the equation~\cite{Franche:2009gk}
\begin{equation}
 \sqrt{2\frac{X}{\Lambda^4}} \frac{\partial p}{\partial X} = A\,,\label{eqofminf}
\end{equation}
with the `non-canonicalness' parameter
\begin{equation}
 A(\phi)=\frac{V'}{3\,H\,\Lambda^2}\,.\label{Adefgen}
\end{equation}
For $A\ll1$ the theory is in its canonical limit, i.e. $p(X,\phi)\simeq X - V(\phi)$ while for $A\gg1$ the theory shows its non-canonical nature, i.e. the terms $X^n$ with $n>1$ dominate the Lagrangian.

For theories with a finite convergence radius $X/\Lambda^4<R$ of $S(X)$, it was shown that a truly non-canonical inflationary solution of eq.~\eqref{eqofminf} with $A\gg1$ exists under the following conditions:
\begin{itemize}
 \item The derivative $\partial_X S(X)$ diverges at the radius of convergence $R$.
 \item The potential is large in units of the cutoff scale, i.e. $V\gg\Lambda^4$ such that the energy density of the potential always dominates that of the kinetic terms~\footnote{Note that the effective field theory description is valid as long as $H<\Lambda$. This generally allows large values of the potential in terms of the cutoff scale since $\frac{H}{\Lambda} \simeq \left( \frac{V}{\Lambda^4} \right)^{1/2}\frac{\Lambda}{M_p}$.}.
\end{itemize}
Note that a finite radius of convergence implies a speed limit $X<R\Lambda^4$. Theories without a speed limit with a $p(X,\phi)$ monotonically increasing in $X$ might lose validity for $X > \Lambda$ as an effective field theory.

The scalar power spectrum $\Delta_s^2$, the tensor power spectrum $\Delta_t^2$, the scalar spectral index $n_s$ and the tensor spectral index $n_t$ can then be calculated via~\cite{Garriga:1999vw,Chen:2006nt}
\begin{align}
\begin{aligned}
 \Delta_s^2(k) &= \left. \frac{1}{8\pi^2} \frac{H^2}{M_p^2} \frac{1}{c_s \epsilon} \right|_{c_s k = a H}\,,\\
 \Delta_t^2(k) &= \left. \frac{2}{\pi^2} \frac{H^2}{M_p^2} \right|_{k = a H}\,,\\
 n_s(k) -1 &= \left.-2 \epsilon - \eta - \kappa\right|_{c_s k = a H}\,,\\
 n_t(k) &= \left. -2 \epsilon\right|_{k = a H}\,.\label{obsgen}
\end{aligned}
\end{align}

In the canonical case, this reduces to
\begin{align}
\begin{aligned}
 \Delta_s^2(k) &= \left. \frac{1}{8\pi^2} \frac{H^2}{M_p^2} \frac{1}{\epsilon} \right|_{k = a H}\,,\\
 \Delta_t^2(k) &= \left. \frac{2}{\pi^2} \frac{H^2}{M_p^2} \right|_{k = a H}\,,\\
 n_s(k) -1 &= \left.-2 \epsilon - \eta \right|_{k = a H}\,,\\
 n_t(k) &= \left. -2 \epsilon\right|_{k = a H}\,.
\end{aligned}
\end{align}

\section{On-shell transformation of inflationary solutions} \label{findcanpot_sec}

In any theory, canonical or non-canonical with scalar field $\chi$, the inflationary solution can be expressed as a function $X_{inf}(\chi)$. We want to obtain the solution $X_{inf}(\phi)$ from a canonically normalized Lagrangian with scalar field $\phi$ and potential $\Vcan(\phi)$. In other words we want to find $\Vcan(\phi)$ such that the slow-roll inflationary solution of the action $P = X - \Vcan(\phi)$, $X_{inf}^{can}(\phi)$, has the same functional form as the inflationary solution $X_{inf}(\chi)$ coming from a general $P(X, \chi)$.  In the following we describe how to construct $\Vcan(\phi)$.

In a canonically normalized theory that allows slow-roll inflation, the equations of motion are approximately
\begin{equation}
 \dot \phi \simeq - \frac{\Vcan'(\phi)}{3 H(\phi)}\,,\qquad H^2(\phi) \simeq \frac{\Vcan(\phi)}{3}\,,\label{eomcan}
\end{equation}
where $'$ denotes the derivative with respect to $\phi$. Using $\dot \phi = - \sqrt{2 X}$ we obtain
\begin{eqnarray}
X & \simeq & \frac{1}{6} \frac{(V_{can}')^2}{V_{can}}\\
 \sqrt{6 X}\,d \phi &=& \frac{1}{\sqrt{\Vcan}} \,d \Vcan\,,\label{dphidV}
\end{eqnarray}
where the first expression is a slow-roll approximation (see e.g. \cite{Franche:2009gk}).  At this point we replace the approximation with an equal sign, since we are looking for a potential which satisfies the slow-roll conditions. 
Now, going on-shell $X=X_{inf}(\chi)$ and hence $d\phi=d\chi$ we can integrate both sides of eq.~\eqref{dphidV} to solve for $\Vcan$:
\begin{align}
\begin{aligned}
&\int_{\phi_0}^\phi \sqrt{6X_{inf}(\chi)} \, d \chi = \int_{{V_0}_{can}}^V \frac{d \Vcan}{\sqrt{\Vcan}}\,,\\
\Rightarrow \quad &\Vcan(\phi) = \left( \sqrt{{V_0}_{can}} + \int_{\phi_0}^\phi \sqrt{\frac{3}{2}\,X_{inf}(\chi)} \, d \chi \right)^2\,,
\end{aligned}
\label{Vtransformed}
\end{align}
with ${V_0}_{can} = \Vcan(\phi_0)$. Eq.~\eqref{Vtransformed} can be seen as an on-shell transformation of the originally possibly non-canonical theory to a canonical theory. It gives us the potential $\Vcan$, whose dynamics described in eq.~\eqref{eomcan} give exactly the same trajectory in phase space as in the original theory. In other words, given an inflationary trajectory in a theory with general kinetic term, we have derived the form of the potential in a theory with canonical kinetic term which will give rise to the same inflationary trajectory. We assume that the kinetic term is canonical and $X = X_{inf}^{can} = X_{inf}^{noncan}= X_{inf}$, and find the corresponding $\Vcan$. This is not a field transformation, since we simply match the inflationary trajectory in two different theories. Hence for any properties regarding the inflationary background solution the fields $\chi$ and $\phi$ are the same while their general dynamics governed respectively by their non-canonical and canonical Lagrangians are 
different. Note that we are free to choose $V_0$ (an integration constant) to satisfy the slow-roll conditions, since we are explicitly looking for a slow-roll solution in a canonical theory with the same inflationary trajectory as that arising from some given non-canonical theory.\footnote{Here we work on-shell, which is to say at the level of the background equation of motion, rather than performing an off-shell field transformation at the level of the action.  Offshell transformations between canonical and non-canonical theories are discussed in Appendix B, where we show that canonical transformations can be used to transform between canonical and noncanonical theories in the case that the theory with noncanonical kinetic term has a dominantly quadratic potential. This method thus appears to be somewhat limited.}

If the original theory is canonical with potential $V(\chi)$, the inflationary trajectory is given by~\cite{Franche:2009gk}
\begin{equation}
 X_{inf}^{can}\,\, = \,\,\frac{(V')^2}{6\,V} \,\, = \,\, \frac{(V_{can}')^2}{6\,V_{can}}\, ,
\end{equation}
such that $\Vcan(\phi) = V(\chi)$.

\section{Comparison of observables}

In this section, we compare the number of efolds $N_e$, the scalar power spectrum $\Delta_s^2$, the tensor power spectrum $\Delta_t^2$ and the scalar spectral index $n_s$ of non-canonical and canonical inflation. We discuss under what conditions these observables will match for a non-canonical theory and a canonical theory whose potential is obtained via eq.~\eqref{Vtransformed} such that it describes the same dynamics as the non-canonical theory.

The natural time measure during inflation is the number of efolds $N_e$ that inflation produces in the time interval $[t_i,t_f]$. It is defined as
\begin{equation}
 N_e = \int_{t_i}^{t_f} H(t) \,dt = \int_{\phi_{N_e}}^{\phi_{end}} \frac{H(\phi)}{\dot \phi} \,d \phi = \int^{\phi_{N_e}}_{\phi_{end}} \left(\frac{V(\phi)}{6 \, X_{inf}(\phi)}\right)^{1/2} \,d \phi\,, \label{Negen}
\end{equation}
where in the last equation we have used $H^2\simeq V/3$~\footnote{{In the following we restrict our analysis to non-canonical theories where the energy density is dominated by the potential, i.e., $H^2 \simeq V/3$.}} and $\dot \phi = -\sqrt{2X}$ on the inflationary trajectory in phase space and $\phi_{end}$ is the field value when inflation ends. In the case of a canonically normalized Lagrangian, this reduces to
\begin{equation}
 N_e = \int_{\phi_{end}}^{\phi_{N_e}} \frac{V(\phi)}{V'(\phi)} \,d\phi = \int_{\phi_{end}}^{\phi_{N_e}} \frac{1}{\sqrt{2\epsilon}} \,d\phi\,.
\end{equation}
The observables are evaluated as functions of the comoving momentum $k$. Due to the fact that the sound speed $c_s$ is generically different from one, the time of horizon crossing for scalar modes is different from the time of horizon crossing for tensor modes. In terms of efolds $N_e$, the different times of horizon crossing are determined via
\begin{eqnarray}
\begin{aligned}
&\text{scalar modes: }\qquad c_s k = a H \,\,&\Leftrightarrow&\,\, \ln k = (N_e -\ln c_s) + \ln H\,,\\
&\text{tensor modes: }\qquad k = a H \,\,&\Leftrightarrow&\,\, \ln k = N_e + \ln H\,.\label{cskaHandkaH}
\end{aligned}
\end{eqnarray}
Hence, the moment of horizon crossing of the scalar modes is earlier than that of the tensor modes and the correction is logarithmic in $c_s$ with $\ln c_s <0$ due to $c_s <1$.
The speed of sound is constrained from the non-observation of non-Gaussianities of the equilateral type to be $c_s \gtrsim 0.1$ such that the correction to horizon crossing is of the order of one efold. We will ignore this correction in the remainder of this section but will discuss its significance in section~\ref{csnot1_sec}. It will turn out that the correction is negligible for $\Delta_s^2$ and $\Delta_t^2$ while it is significant for $n_s$.

\subsection{Theories with a speed limit}	\label{theoriesspeedlimit_sec}

Let us examine under which conditions the observables of non-canonical inflation and canonical inflation obtained as a function of $N_e$, as discussed in section~\ref{findcanpot_sec}, will agree. Let us make two assumptions:
\begin{itemize}
 \item The non-canonical theory has a canonical branch where $\Vcan \simeq V$.
 \item The non-canonical theory has a speed limit $R$ such that $X_{inf} \simeq \Lambda^4 R$ for $A\gg 1$.
\end{itemize}

We can perform the integration in eq.~\eqref{Vtransformed} analytically and obtain
\begin{equation}
\Vcan(\phi) = \frac{3}{2}R\,\Lambda^4(\phi-C)^2\,,
\end{equation}
with a constant $C$ for the canonical potential in the limit for $A\gg 1$. This implies
\begin{equation}
 \epsilon_{can} = \frac{1}{2} \left(\frac{\Vcan'}{\Vcan} \right)^{2} = \frac{3 R \Lambda^4}{\Vcan(\phi)}\,.\label{epscanAlarge}
\end{equation}
It was shown in~\cite{Franche:2009gk} that the first slow-roll parameter becomes
\begin{equation}
 \epsilon = \sqrt{2R} \,\frac{\epsilon_{V}}{A}
\end{equation}
for $A\gg 1$. Using the definition of $A$, eq.~\eqref{Adefgen}, and eq.~\eqref{epscanAlarge}, the agreement of $\Delta_s^2$ and $\Delta_t^2$ as a function of $\phi$ can be phrased as conditions on the potentials and the speed of sound, i.e.
\begin{equation}
 \Vcan \simeq V \qquad \text{and} \qquad c_s = \frac{\sqrt{2R}}{A} \qquad \text{for } A\gg1\,. \label{condcsVagree}
\end{equation}
Note that the first condition in eq.~\eqref{condcsVagree} is trivially satisfied in the canonical limit $A\ll1$. In the non-canonical limit $A\gg1$, the derivative $V'$ will generically have large values while $\Vcan'$ has to be small in order to support slow-roll inflation. Thus, at some value $A^*$ in the $A\gg1$ limit, $V$ and $\Vcan$ will not agree anymore. However, there can be an intermediate regime $A\in[1,A^*]$ with $\Vcan \simeq V$ and $\Vcan'\ll V'$. This intermediate regime can even serve to describe the complete phenomenologically interesting region if $c_s(A^*)<0.1$, such that only the region $A>A^*$ is excluded due to non-observation of equilateral non-Gaussianities.

The first condition in eq.~\eqref{condcsVagree} implies an agreement as a function of $N_e$ as well since according to eq.~\eqref{Negen}
\begin{align}
\begin{aligned}
 &\text{canonical: } \qquad N_e=\int_{\phi_{end}}^{\phi_{N_e}} \frac{1}{\sqrt{2\epsilon_{can}}} \,d\phi = \int_{\phi_{end}}^{\phi_{N_e}} \left(\frac{\Vcan(\phi)}{6R\Lambda^4}\right)^{1/2} \,d\phi\,,\\
 &\text{non-canonical: } \qquad N_e = \int^{\phi_{N_e}}_{\phi_{end}} \left(\frac{V(\phi)}{6 \, X_{inf}(\phi)}\right)^{1/2} \,d \phi = \int^{\phi_{N_e}}_{\phi_{end}} \left(\frac{V(\phi)}{6R\Lambda^4}\right)^{1/2} \,d \phi\,.
\end{aligned}
\end{align}

As far as the spectral indices $n_s$ and $n_t$ are concerned we do not find agreement in the limit $A\gg 1$ since
\begin{align}
\begin{aligned}
 \text{canonical: } \qquad &n_s-1=-6\epsilon_{can} + 2\eta_{can} = -\frac{12R\Lambda^4}{\Vcan}\,,\\
 &n_t=-2\epsilon_{can}=-\frac{6R\Lambda^4}{\Vcan}\,,\\
 \text{non-canonical: } \qquad &n_s-1=-2 \epsilon - \eta - \kappa = \frac{\sqrt{2R}}{A}(-6 \epsilon_{V}+2\eta_{V})-\kappa\,,\\
 &n_t=-2\epsilon=-\frac{2\sqrt{2R}}{A}\epsilon_{V}\,,\\
\end{aligned}
\end{align}
using $\eta=\sqrt{2R}/A(4\epsilon_{V}-2\eta_{V})$ as was shown in~\cite{Franche:2009gk}. However, this does not exclude an agreement in an intermediate region $A\gtrsim1$. Furthermore, the scalar spectral index $n_s$ receives significant corrections from the fact that $c_s<1$ in non-canonical theories. This can improve the agreement, as we will show in section~\ref{csnot1_sec}.

Let us now investigate with some examples when the second condition in eq.~\eqref{condcsVagree} on the speed of sound $c_s$ can be fulfilled. First, we note that using eq.~\eqref{eqofminf} the speed of sound can be expressed as
\begin{equation}
 c_s^2(A) = \frac{A\, \partial X_{inf} / \partial A}{2 X_{inf}}\,.\label{csXinfA}
\end{equation}
Hence, we need to know the functional dependence $X_{inf}(A)$ in order to decide whether the observables $\Delta_s^2$ and $\Delta_t^2$ of the canonical and non-canonical theory agree. For $p(X,\phi) = \Lambda^4 S(X)-V(\phi)$ as defined in eq.~\eqref{psum} this dependence is determined by the identity
\begin{equation}
 2\frac{X}{\Lambda^4}\left(\sum_{n\geq 0} (n+1)\, c_n \left(\frac{X}{\Lambda^4} \right)^n  \right)^2 = A^2\,,\label{XinfAinv}
\end{equation}
using the algebraic equation for the inflationary solution, eq.~\eqref{eqofminf}. To obtain $X_{inf}(A)$ we have to invert eq.~\eqref{XinfAinv}, which is impossible for most general coefficients $c_n$. However, we will discuss some closed form expressions for $p(X,\phi)$ in the following.

Consider the class of non-canonical Lagrangians defined by
\begin{equation}
 p(X,\phi)= \Lambda^4 \left[1-\left(1-\frac{1}{a}\,\frac{X}{\Lambda^4} \right)^a \right] - V(\phi)\,,\label{classpa}
\end{equation}
with $0<a<1$ such that $\partial p / \partial X$ diverges at the radius of convergence $R_a = a$. This class of non-canonical Lagrangians includes the DBI action via the case $a=1/2$, i.e.
\begin{equation}
 p(X,\phi)= \Lambda^4 \left[1-\left(1-2\,\frac{X}{\Lambda^4} \right)^{1/2} \right] - V(\phi)\,.
\end{equation}

Squaring the equation for the inflationary solution, eq.~\eqref{eqofminf} becomes
\begin{equation}
 2 \frac{X}{\Lambda^4} = A^2 \left(1- \frac{1}{a}\,\frac{X}{\Lambda^4} \right)^{2-2a}\,.\label{infsola}
\end{equation}
If $2-2a$ is not an integer one has to exponentiate with the denominator of $2-2a$ to solve for $X_{inf}(A)$. In fact the only value of $0<a<1$ for which $2-2a$ is an integer is $a=1/2$, i.e. the DBI case, with solution
\begin{equation}
 X_{inf} = \frac{\Lambda^4}{2}\,\frac{A^2}{1+A^2}\,.\label{XinfDBI}
\end{equation}
For all $a\neq 1/2$, $X_{inf}$ will be some function of $A^n$ with integer $n>2$. For instance, for $a=3/4$ we have to square eq.~\eqref{infsola} to obtain the solution
\begin{equation}
 X_{inf} = \frac{\Lambda^4}{6}\,A^4 \left(\sqrt{1+\frac{9}{A^4}} - 1 \right)\,.
\end{equation}

Note that for $X_{inf}(A^n)$, $c_s^2$ is also a function of $A^n$ since
\begin{equation}
 c_s^2 = \frac{n A^n \,X_{inf}'(A^n)}{2X_{inf}(A^n)}\,,
\end{equation}
where $'$ denotes the derivative with respect to $A^n$. Hence, the dominating term in $c_s$ will be of the order
\begin{equation}
 c_s^2 \sim \frac{1}{A^n}
\end{equation}
up to an $\mathcal{O}(1)$ coefficient. For the DBI case, we find
\begin{equation}
 c_s^2 = \frac{1}{1+A^2} \simeq \frac{1}{A^2} \qquad \text{for } A\gg1\,,\label{csDBI}
\end{equation}
which fulfills the criterion eq.~\eqref{condcsVagree} on $c_s$ for the agreement of the observables ($R=1/2$). However, this is the only member of the class of non-canonical theories defined by eq.~\eqref{classpa} where the observables can agree since 
$c_s^2 \sim 1/A^n$ with $n>2$ for all other values of $a$ such that the condition on $c_s$ in eq.~\eqref{condcsVagree} cannot be satisfied. For example, for $a=3/4$ we find
\begin{equation}
 c_s^2 = 1- \frac{1}{\sqrt{1+9\,A^{-4}}} \simeq \frac{9}{2}\, \frac{1}{A^4} \qquad \text{for } A\gg1\,,
\end{equation}

There are of course plenty of other models apart from those defined in eq.~\eqref{classpa} that fulfill the conditions of a canonical branch and a speed limit. The question of whether there could be other examples than DBI where the conditions on the potential and speed of sound eq.~\eqref{condcsVagree} for an agreement of $\Delta_s^2$ and $\Delta_t^2$ are fulfilled is hard to answer in full generality. Consider for example the class of functions
\begin{equation}
 p(X,\phi) = X\left[1- a \left(\frac{X}{\Lambda^4} \right)^{b} \right]^{c}\,.
\end{equation}
For $a=4$, $b=4$ and $c=1/2$ we numerically find a solution $X_{inf}(A)$ of the equations of motion eq.~\eqref{eqofminf} with
\begin{equation}
 c_s^2 \simeq \frac{\sqrt{2}}{A^2} = \frac{2 R}{A^2} \qquad \text{ for } A\gg1\,,
\end{equation}
such that the second condition in eq.~\eqref{condcsVagree} on the speed of sound is fulfilled. However, this solution suffers from the absence of a canonical limit $X_{inf} \sim A^2$ for $A<1$ and a violation of the null-energy condition $\partial p/ \partial X > 0$. Due to the lack of other working examples where the agreement conditions eq.~\eqref{condcsVagree} are matched, we suspect that the description in terms of a canonical theory may be special to the DBI case. We will study this case more explicitly in the following section. We note at this point that the matching of the background equation of motion does not necessarily mean that fluctuations around this background in the two different theories should match. One should thus not expect agreement of the inflationary observables in general, even if the inflationary trajectory is the same. This makes the agreement in the DBI case all the more remarkable. \footnote{We thank Bret Underwood for discussions on this point.}

\subsection{DBI inflation with an inflection point potential} \label{DBIinflectionexample_sec}

We now want to give an example of our general considerations in section~\ref{theoriesspeedlimit_sec}. We consider the DBI action together with an inflection point potential:
\begin{equation}
 p(X,\phi)=-\frac{1}{f(\phi)} \left(\sqrt{1- 2f(\phi)X} -1 \right) - V(\phi)\,,\label{DBIinflp}
\end{equation}
with
\begin{equation}
 V(\phi)= V_0 + \lambda (\phi-\phi_0)+\beta (\phi-\phi_0)^3\,.\label{DBIinflpointpotential}
\end{equation}
We fix the parameters of this theory to be
\begin{equation}
V_0 = 3.7 \cdot 10^{-16}\,,\quad \lambda = 1.13\cdot 10^{-20}\,,\quad \beta = 1.09\cdot 10^{-15}\,,\quad \phi_0 = 0.01\,,\quad f = 1.6\cdot 10^{21} \,.\label{DBIinflparam}
\end{equation}
These are the values that were considered in~\cite{Franche:2009gk}. In particular, the field-dependent warp factor has been set to a constant $f=\Lambda^{-4}$ which is justified if the range of field values that $\phi$ travels during inflation is small. The parameters in eq.~\eqref{DBIinflparam} have been chosen such that for a canonical kinetic term $p(X,\phi)=X-V$ the amplitude of the scalar fluctuations and the spectral index agree with observations, i.e. $\Delta_s^2 = 2.41 \cdot 10^{-9}$ and $n_s = 0.961$.

\begin{figure}[h!]
\centering
\includegraphics[width= 0.5\linewidth]{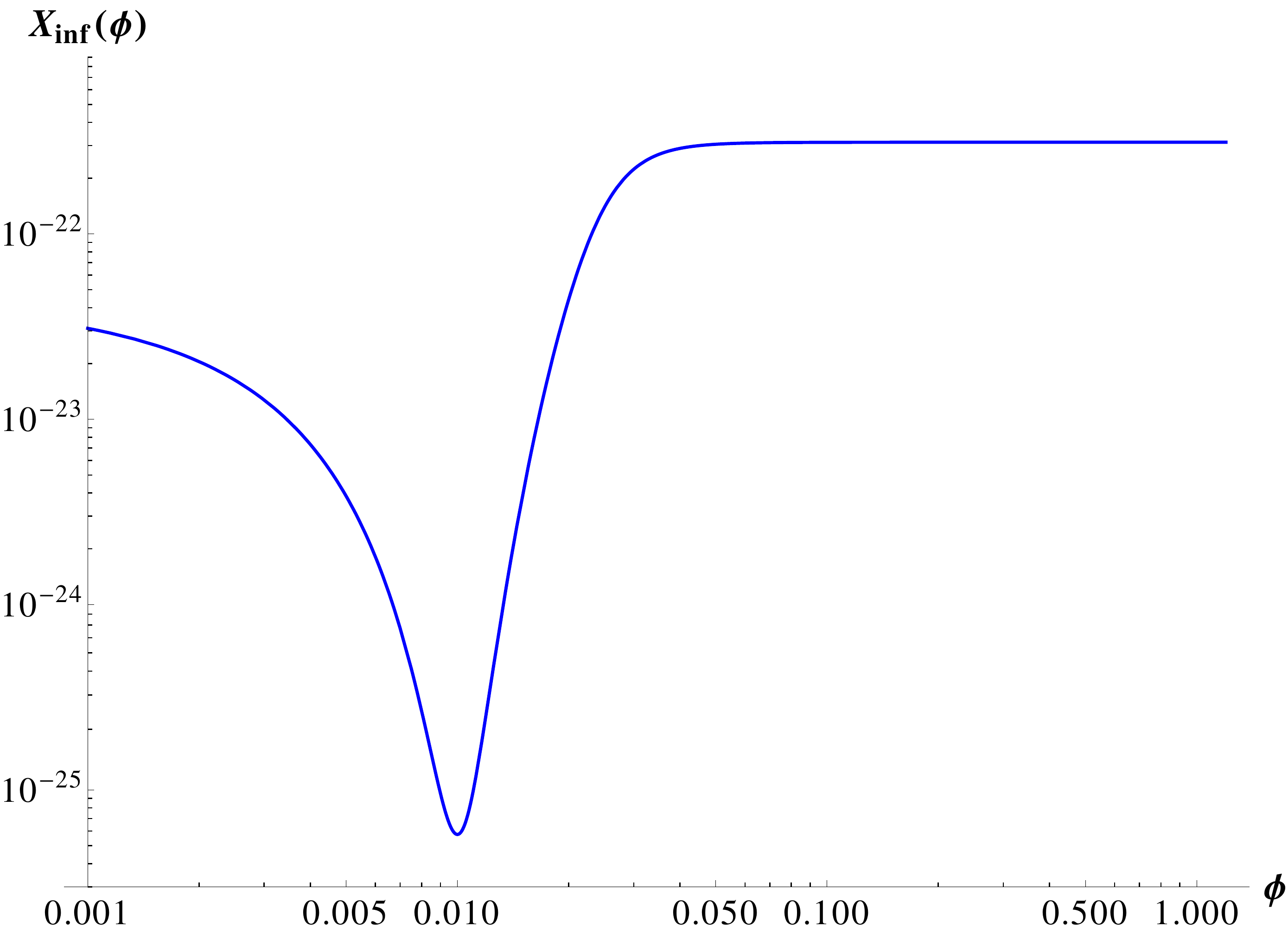}
\caption{The phase space trajectory $X_{inf}(\phi)$ for the DBI action eq.~\eqref{DBIinflp} with the numerical values given in eq.~\eqref{DBIinflparam}. For large $\phi$ the trajectory approaches the limit $(2f)^{-1}$, see eq.~\eqref{XinfDBI}.}
\label{Xinfaplot_DBI_fig}
\end{figure}

\begin{figure}[h!]
\vskip -3mm
\centering
\includegraphics[width= 0.49\linewidth]{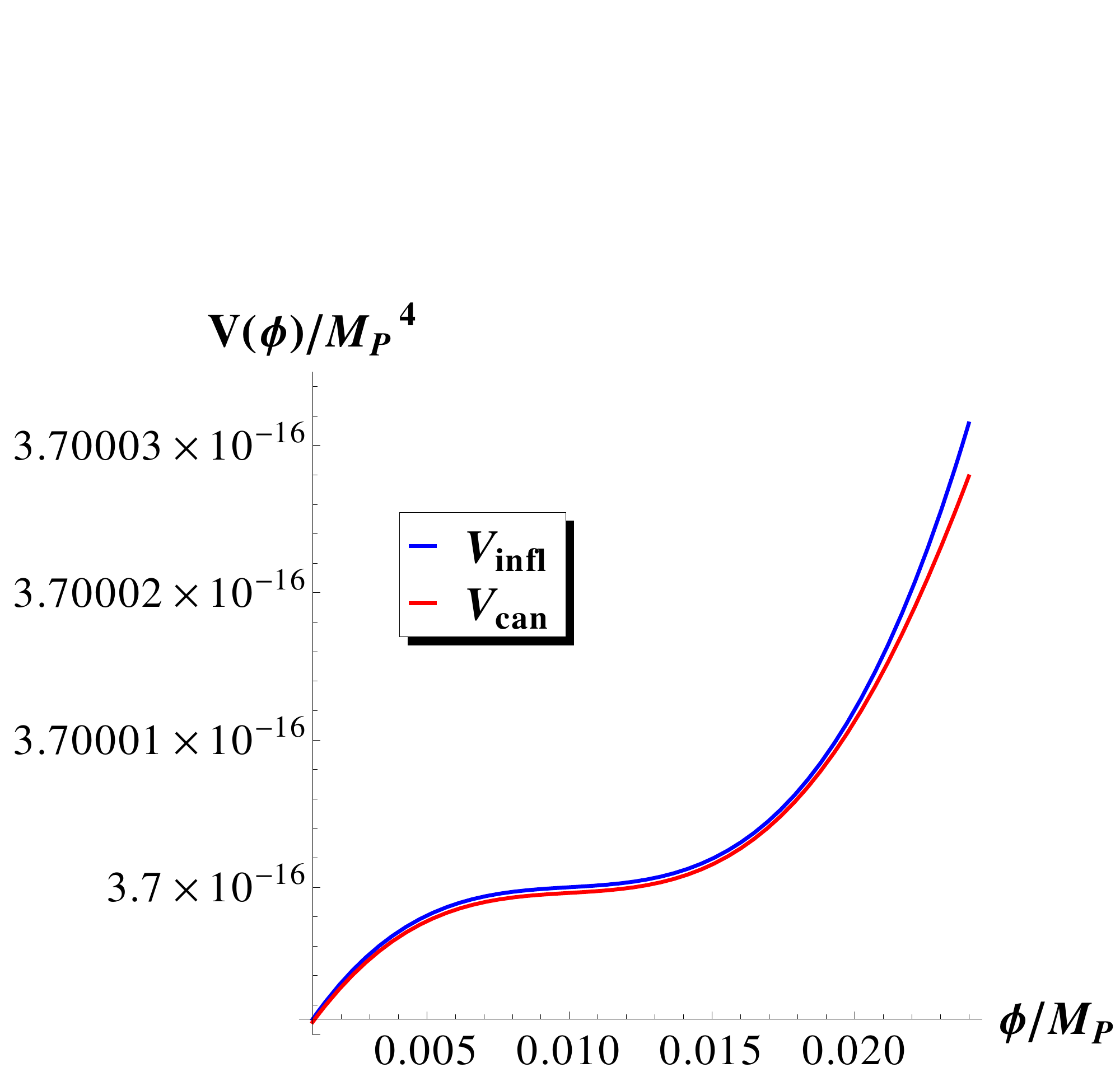}
\includegraphics[width= 0.49\linewidth]{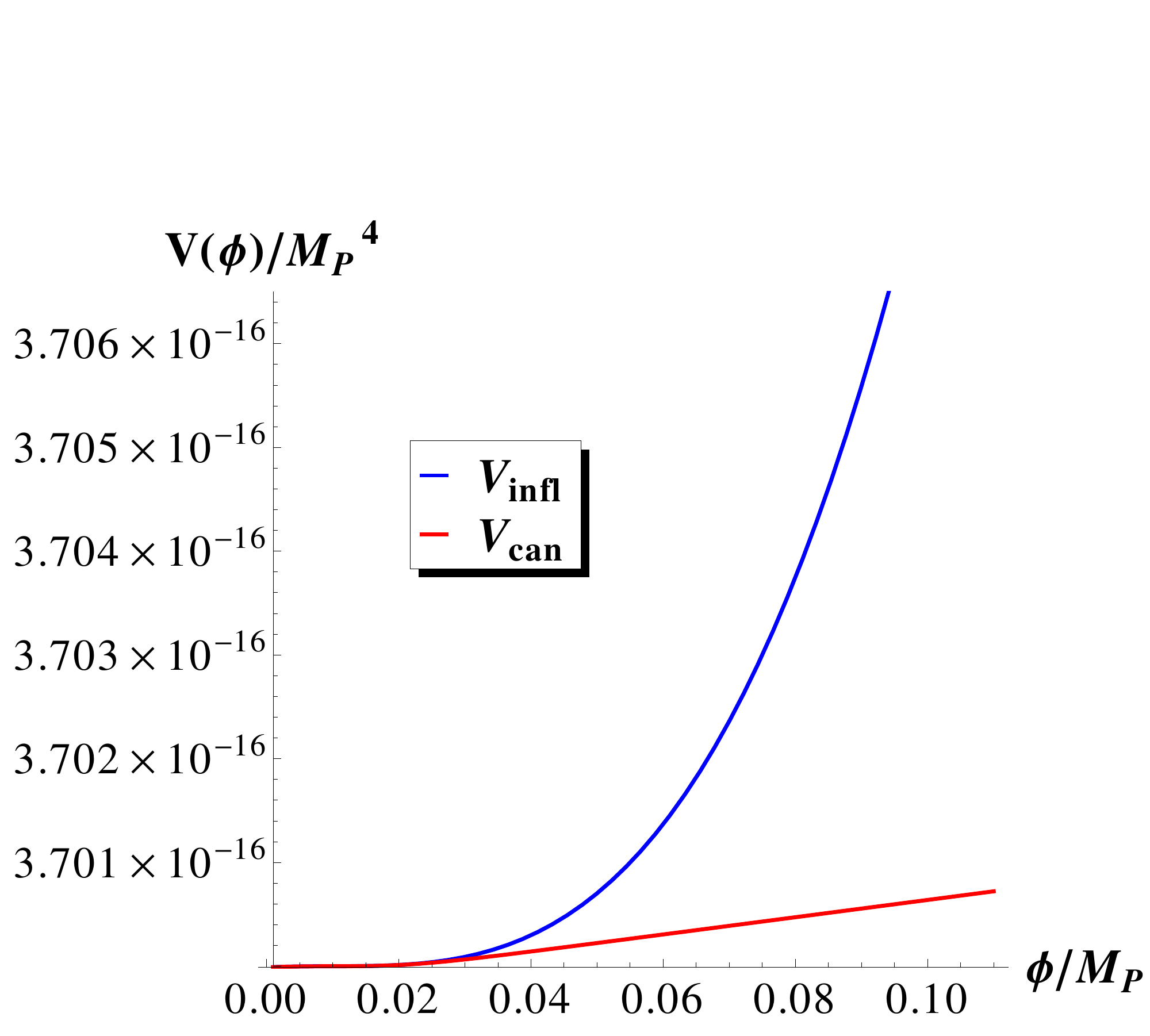}
\caption{Comparison of the inflection point potential $V\equiv V_{infl}$ of eq.~\eqref{DBIinflpointpotential} and the potential of the canonical theory $\Vcan(\phi)$ obtained via eq.~\eqref{Vtransformed} for $\phi \in [0,0.025]$ (left) and $\phi \in [0,0.12]$ (right).}
\label{VcanvsVinf_fig}
\end{figure}

\begin{figure}[h!]
\centering
\includegraphics[width= \linewidth]{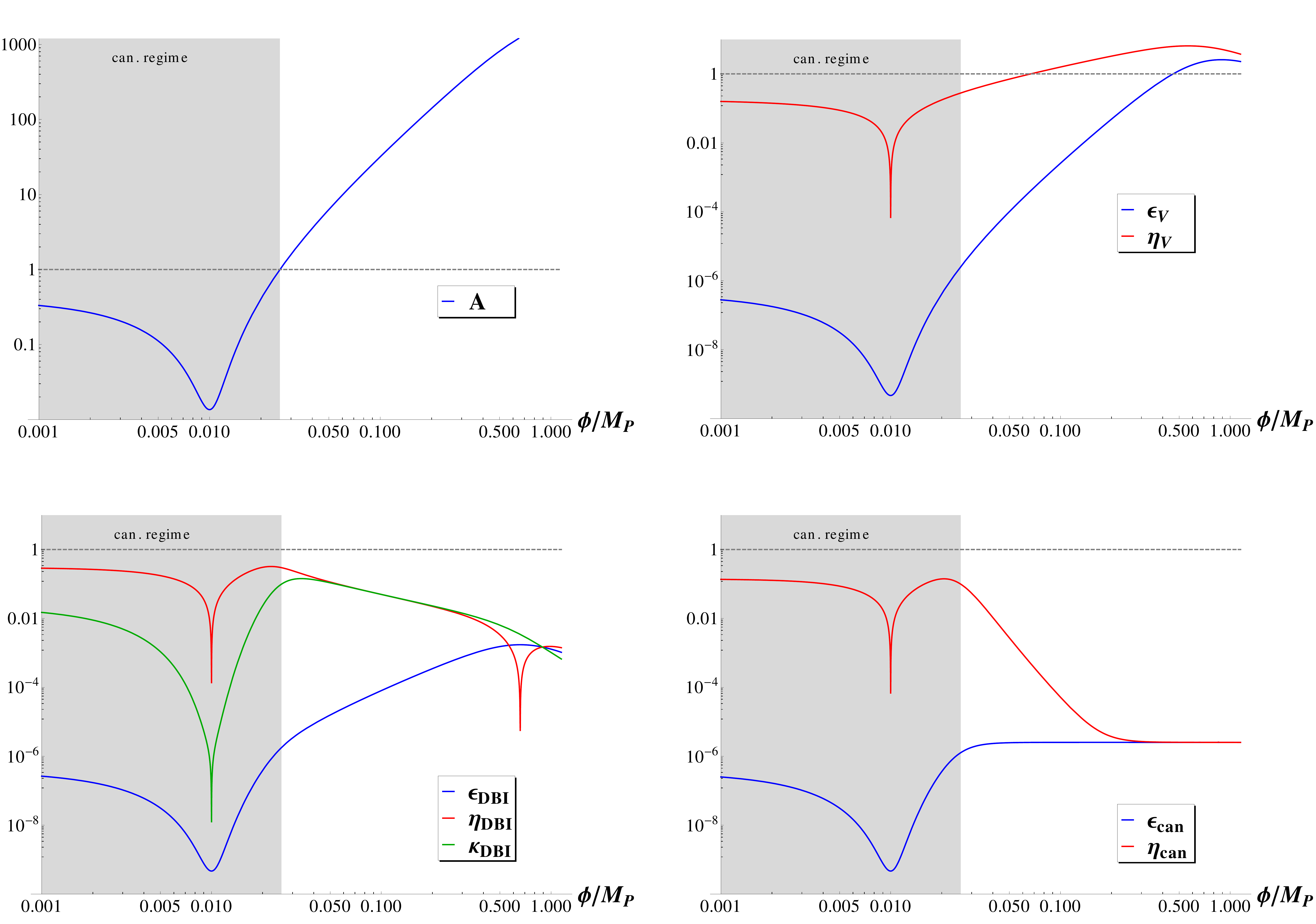}
\caption{The `non-canonicalness' parameter $A$ (top left), the parameters $\epsilon_V$ and $\eta_V$ (top right) and the generalized slow-roll parameters $\epsilon$, $\eta$ and $\kappa$ (bottom left) for the DBI action eq.~\eqref{DBIinflp} with the numerical values of the parameters given in eq.~\eqref{DBIinflparam}. Also the slow roll parameters $\epsilon_{can}$ and $\eta_{can}$ of the canonical theory are shown (bottom right).}
\label{epsiloncompare_fig}
\end{figure}

Let us first see when eq.~\eqref{DBIinflp} is in the non-canonical regime by evaluating the `non-canonicalness' parameter $A$.
We find that for $\phi \lesssim 0.025$ we are in the canonical regime $A\leq1$, while for $\phi \gtrsim 0.025$ we enter the non-canonical regime $A>1$, see Figure~\ref{epsiloncompare_fig}.
The phase space trajectory (see also Figure~\ref{Xinfaplot_DBI_fig}) for eq.~\eqref{DBIinflp} is determined by eq.~\eqref{XinfDBI}.
This determines the potential $\Vcan(\phi)$ that resembles the trajectory from a canonical kinetic term via eq.~\eqref{Vtransformed}. We perform the integration numerically and show $\Vcan(\phi)$ compared to the original inflection point potential $V(\phi)$ in Figure~\ref{VcanvsVinf_fig}.
We see that, as expected, $\Vcan$ agrees with $V$ in the canonical regime while it is flatter than $V$ in the non-canonical regime. To see that $\Vcan$ actually supports slow-roll inflation we check $\epsilon$ and $\eta$ as functions of $\phi$ in Figure~\ref{epsiloncompare_fig}.

\subsubsection*{Comparison of observables}

\begin{figure}[h!]
\centering
\includegraphics[width= \linewidth]{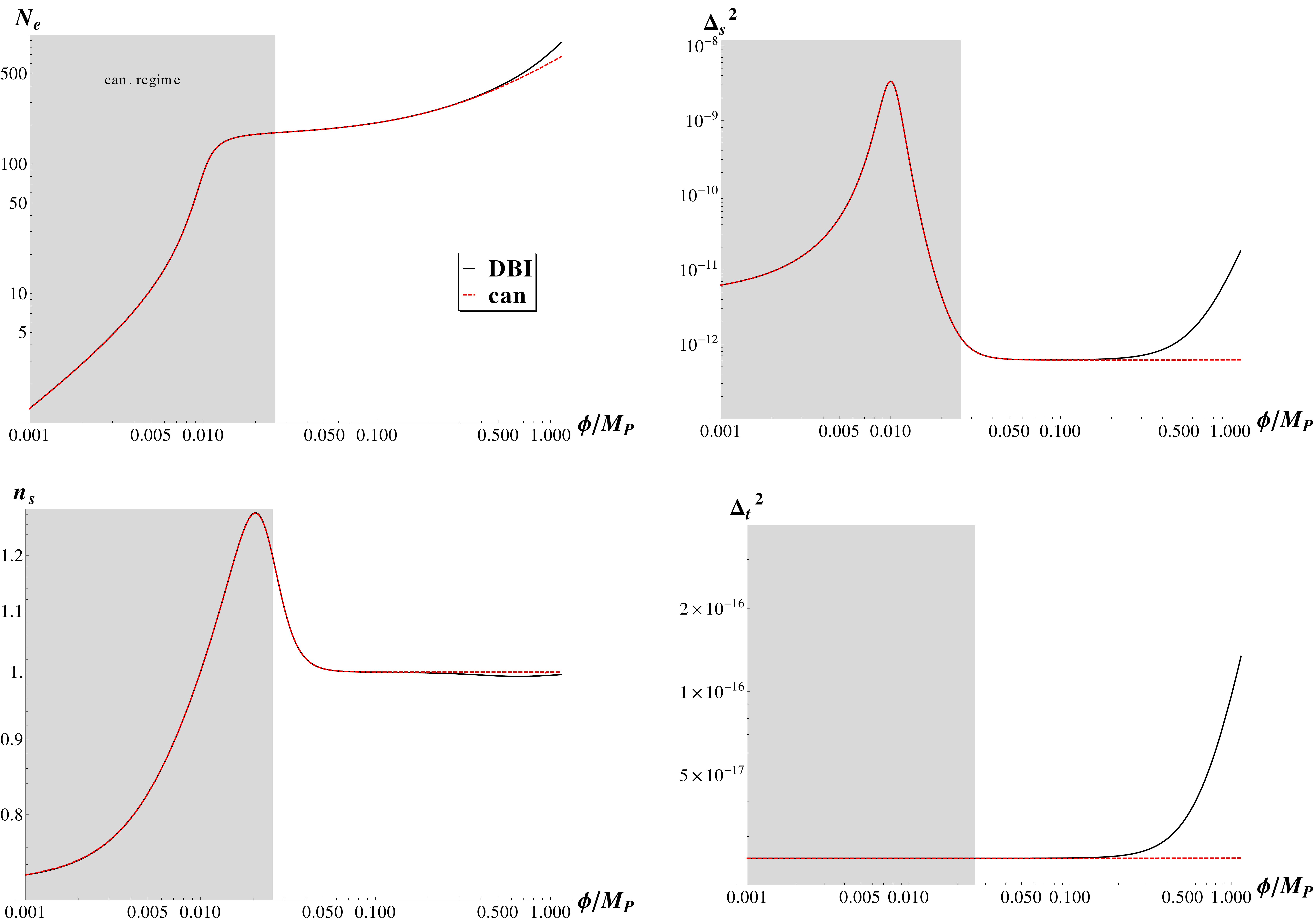}
\caption{Comparison of the observables $\Delta_s^2$ (top right), $n_s$ (bottom left) and $\Delta_t^2$ (bottom right) of the non-canonical DBI and the transformed canonical theory. Since the number of efolds (top left) of the two theories agrees as a function of $\phi$, the agreement of the observables as a function of $\phi$ can be read as an agreement as a function of $N_e$.}
\label{Necompare_fig}
\end{figure}

We compare the observables of the canonical and non-canonical theory in Figure~\ref{Necompare_fig}. The agreement in $\Delta_s$ and $\Delta_t$ at the level of $\sim 1 \%$ is up to values $\phi<0.2$ which is roughly one order of magnitude more than the value of $\phi$ where the non-canonical regime begins. So as discussed after eq.~\eqref{condcsVagree} there is indeed an intermediate regime where the observables agree even though $\Vcan$ is much flatter than $V$. Furthermore, since $c_s < 0.1$ for $\phi > 0.06$ the phenomenologically viable region is included in this intermediate regime.
The agreement of $n_s-1$ of the two theories as functions of $N_e$ holds only up to $\phi \leq 0.05$, see Figure~\ref{nscscorr_fig}. However, there are important corrections to $n_s-1$ induced by the fact that the speed of sound $c_s$ in the non-canonical theory is smaller than one. We will discuss these corrections in detail in section~\ref{csnot1_sec}.

Note that there is an additional upper bound on $c_s$ which has to be fulfilled in order to treat the inflationary quantum fluctuations perturbatively~\cite{Cheung:2007st,Leblond:2008gg,Shandera:2008ai}. If the speed of sound becomes too small the perturbations become strongly coupled and in particular the expressions for the inflationary observables eq.~\eqref{obsgen} are not valid. For DBI this can be expressed as a bound on the `non-canonicalness' parameter~\cite{Franche:2009gk}
\begin{equation}
 A < \left( \frac{3\,\epsilon}{V}\right)^{1/5}\,.
\end{equation}
For our numerical example, this implies $A < \mathcal{O}(100)$ and hence $\phi \lesssim 0.2$. Note that this is exactly the region where we find agreement between the non-canonical and transformed canonical theory.

We can prove the agreement of $\Delta_s^2$ and $\Delta_t^2$ in the whole intermediate region (note that in section~\ref{theoriesspeedlimit_sec} this was shown only in the limit $A\gg1$). Using the exact expression for the speed of sound in eq.~\eqref{csDBI}, together with eq.~\eqref{epscanAlarge}, we find
\begin{equation}
 \epsilon_{can} = \frac{3\Lambda^4}{2 \Vcan} \frac{A^2}{1+A^2}\, ,
\end{equation}
and the exact expression for $\epsilon$ that was found in~\cite{Franche:2009gk} is
\begin{equation}
 \epsilon = \frac{3}{2}\, \frac{A^2}{1+A^2}\, \frac{1}{1+\frac{V/\Lambda^4-1}{\sqrt{1+A^2}}}\,.
\end{equation}
Now the condition $c_s \epsilon = \epsilon_{can}$ can be rephrased as
\begin{equation}
 \frac{V}{\Lambda^4} + \sqrt{1+A^2} - 1 = \frac{\Vcan}{\Lambda^4}\,.\label{agreementDBIanalytic}
\end{equation}
This condition will be fulfilled to very large $A$ for $V\simeq \Vcan$, since $V\gg \Lambda^4$ as we demanded at the beginning of section~\ref{theoriesspeedlimit_sec}. For instance, in the numerical example described in eq.~\eqref{DBIinflparam} we have $V/\Lambda^4 \simeq 10^5$ such that eq.~\eqref{agreementDBIanalytic} would hold up to $A\lesssim 10^4$
assuming the condition $V\simeq \Vcan$ is not violated before $A$ reaches this value.

The agreement works out as well for the DBI action with a Coulomb type potential
\begin{equation}
 V(\phi) = V_0 - \frac{T}{(\phi+\phi_0)^n}\,,
\end{equation}
instead of an inflection point potential. The non-canonical regime is accessed for $\phi < \phi_0$ while the canonical regime is given by $\phi > \phi_0$. Hence, the agreement with the transformed canonical theory is trivially found for $\phi < \phi_0$ and extends to the non-canonical regime until the condition $V \simeq \Vcan$ is violated.

\subsubsection*{Consistency relation}
Canonical and non-canonical theories are usually assumed to be distinguishable, not only because of the possibility of equilateral-type non-gaussianity in the latter, but because of the consistency relation relating $r = \Delta_t^2/\Delta_s^2$ the tensor-to-scalar ratio and $n_t$ the tensor spectral index. The relation in the noncanonical case has an additional factor of $c_s$ \cite{Garriga:1999vw}:
\begin{eqnarray*}
r_{Can} & = & - 8 n_t;\\
r_{NC} & = & - 8 c_s n_t.
\end{eqnarray*}
Because of the appearance of $c_s$, a sufficiently precise measurement of the ratio $r/n_T$ would therefore resolve the degeneracy we have found. However, we currently have no bound on $n_t$ and only an upper bound on $r$: $r< 0.12$ \cite{Ade:2013uln}. With the current state of observational bounds, these models remain indistinguishable at the 2-point function level. 
\section{Corrections from $c_s < 1$} \label{csnot1_sec}

As we discussed in eq.~\eqref{cskaHandkaH}, the observables have to be evaluated as functions of the comoving momentum $k$ which implies different times of horizon crossing for scalar and tensor modes respectively. Assuming $H_{can} \simeq H_{non-can}$ which actually follows from the condition $\Vcan \simeq V$, an agreement of tensor observables $T$ as functions of $\ln k$ is equivalent to
\begin{equation}
 T_{can} (N_e) = T_{non-can} (N_e)\,,
\end{equation}
having used $N_e = \ln k - \ln H$.

For scalar observables $S$ however, we have to take into account that $N_e - \ln c_s = \ln k - \ln H$ in the non-canonical theory while $N_e = \ln k - \ln H$ in the canonical theory. Hence, we have to check for the equality
\begin{equation}
 S_{can} (N_e) = S_{non-can} (N_e - \ln c_s)\,.
\end{equation}
Since the non-observation of equilateral non-Gaussianities implies $|\ln c_s| \ll N_e^t$, it is sufficient to expand $S_{non-can}$ to first order in $\ln c_s$, i.e.
\begin{equation}
 S_{non-can} (N_e - \ln c_s) \simeq S_{non-can} (N_e) - S_{non-can}' (N_e)\, \ln c_s\,.
\end{equation}
In the following, we discuss this expansion for the scalar power spectrum $\Delta_s^2$ and the scalar spectral index $n_s$.

Using the definition of $\Delta_s^2$ in eq.~\eqref{obsgen}, we find
\begin{align}
\begin{aligned}
\frac{\partial \Delta_s^2}{\partial N_e} &=  \Delta_s^2 \,\frac{\partial \ln \Delta_s^2}{\partial N_e} = \Delta_s^2 \left(2\frac{\partial \ln H}{\partial N_e} - \frac{\partial \ln \epsilon}{\partial N_e} - \frac{\partial \ln c_s}{\partial N_e}\right)\,,\\
&= \Delta_s^2 \cdot \left(-2\epsilon - \eta - \kappa \right) = \Delta_s^2 \cdot (n_s - 1)\,,
\end{aligned}
\end{align}
having used
\begin{equation}
 \epsilon = - \frac{\partial \ln H}{\partial N_e}\,,\qquad \eta= \frac{\partial \ln \epsilon}{\partial N_e} \,,\qquad \kappa = \frac{\partial \ln c_s}{\partial N_e}\,.
\end{equation}
This implies
\begin{equation}
 \Delta_s^2(N_e - \ln c_s) \simeq \Delta_s^2(N_e) \left[1- (n_s-1)\ln c_s \right]\,.
\end{equation}
Hence the correction that is induced by $\ln c_s$ is suppressed by the slow-roll parameters and we can approximate $\Delta_s^2(N_e - \ln c_s) \simeq \Delta_s^2(N_e)$.
\begin{figure}[t!]
\centering
\includegraphics[width= 0.8\linewidth]{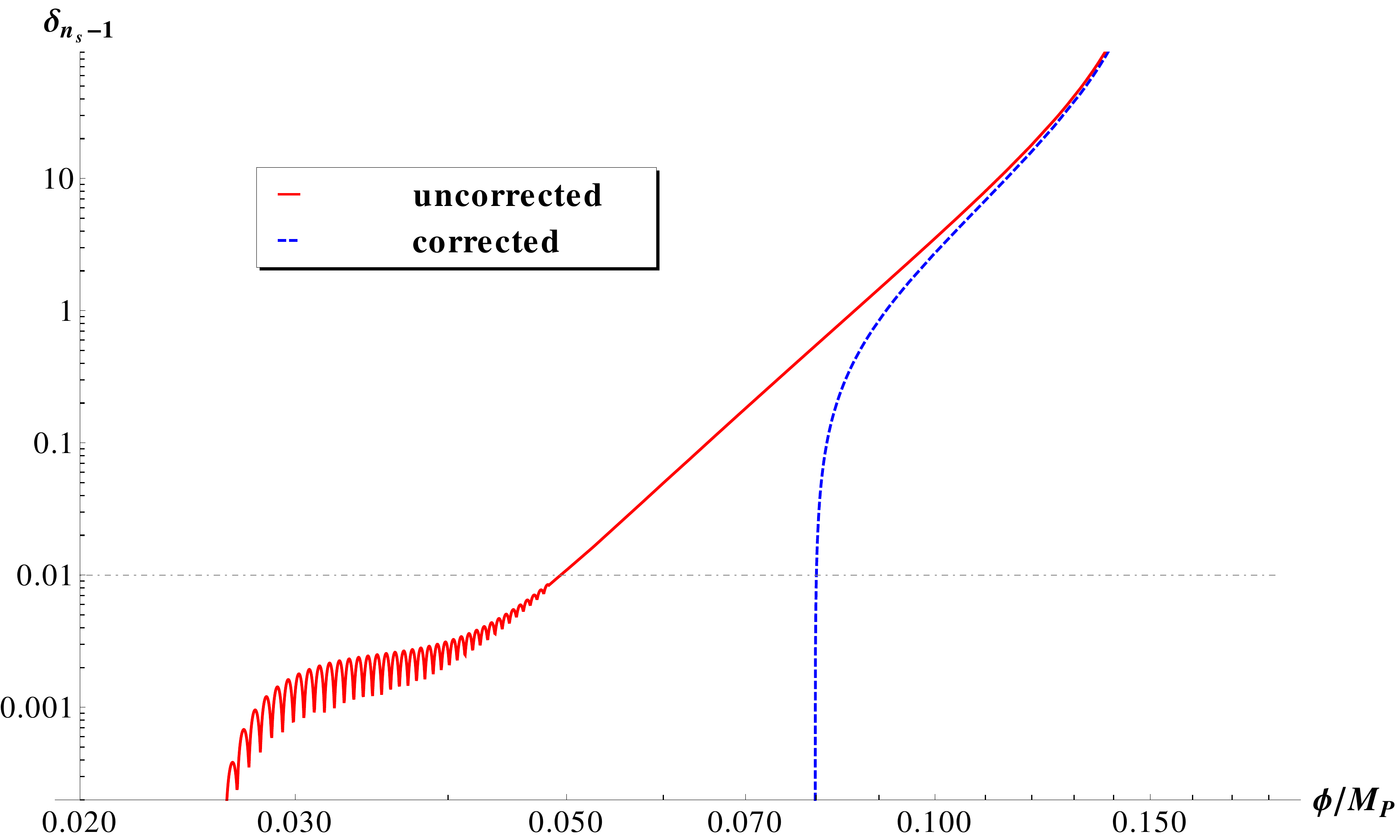}
\caption{The agreement $\delta_{n_s-1}$ of the canonical and non-canonical theory in $n_s$ defined in eq.~\eqref{nscscorr} as a function of $\phi$ with and without $c_s$ corrections. The fluctuations in the uncorrected $\delta_{n_s-1}$ for small $\phi$ are due to numerical inaccuracies when obtaining $\Vcan$ via numerical integration.}
\label{nscscorr_fig}
\end{figure}

For the spectral index $n_s$ the corrections induced by $\ln c_s$ are significant. Using the definition of $n_s$ in eq.~\eqref{obsgen} we find to first order in the slow-roll parameters
\begin{equation}
 \frac{\partial n_s}{\partial N_e} = - \eta \frac{\partial \ln \eta}{\partial N_e}- \kappa \frac{\partial \ln \kappa}{\partial N_e}\,,
\end{equation}
which implies
\begin{equation}
 n_s(N_e - \ln c_s) -1 \simeq -2 \epsilon - \eta - \kappa + \left( \eta \frac{\partial \ln \eta}{\partial N_e} + \kappa \frac{\partial \ln \kappa}{\partial N_e} \right) \ln c_s\,.\label{nscscorr}
\end{equation}
Note that $\partial \ln \eta / \partial N_e$ corresponds to third derivative terms of the potential $V$, while the $\partial \ln \kappa / \partial N_e$ term corresponds to second derivative terms of the speed of sound $c_s$.

We show in Figure~\ref{nscscorr_fig} numerically that the agreement in
\begin{equation}
 \delta_{n_s-1} \equiv \frac{(n_s-1)_{can} - (n_s-1)_{non-can}}{(n_s-1)_{can}}\,,
\end{equation}
for the DBI example considered in section~\ref{DBIinflectionexample_sec} improves if one takes the corrections described in eq.~\eqref{nscscorr} into account. We find that the regime where $n_s-1$ of the canonical and non-canonical theory agree at the level of $1\%$ is increased from $\phi \leq 0.05$ to $\phi \leq 0.08$. Consequently, the phenomenologically interesting region where $c_s > 0.1$ given by $\phi \leq 0.06$ is included due to the inclusion of this correction. 

\section{Conclusions}\label{sec:conc}

Cosmological inflation generates almost scale-invariant power spectra of super-horizon size curvature perturbations and tensor mode perturbations with Gaussian distribution. They can be described at the Gaussian level by just three observational quantities: The overall normalization $\Delta_s^2$ of the curvature perturbation power spectrum, its spectral tilt $n_s$, describing the (small) deviations from scale invariance expected in most models of inflation, and the fractional power $r$ in tensor modes.

In this context it is important to understand the structure of this very large model space, and look for degeneracies between large classes of inflationary models with respect to the three observable quantities. We have restricted our attention here to single-field models of inflation which partition into two large classes: models with a canonically normalized kinetic term $\frac12 (\partial_\mu\phi)^2$, and so-called non-canonical inflation models with Lagrangian ${\cal L}=p\left((\partial_\mu\phi)^2,\phi\right)$.

We have explored the degeneracies between canonical and non-canonical models of inflation with respect to the three observational quantities describing their predicted power spectra at the Gaussian level. While formally non-canonical 2-derivative models of the form ${\cal L}=f(\phi)(\partial_\mu\phi)^2-V(\phi)$ can be always transformed off-shell by a local field redefinition into a canonical model with a transformed scalar potential, this question is rather non-trivial in the presence of higher-power kinetic terms. We have elucidated the method of canonical transformations for transforming noncanonical kinetic terms into canonical kinetic terms which, even in $0+1$D, appears to be limited to the case where the noncanonical theory has a quadratic potential, see Appendix \ref{sec:AppB}. 

As the inflationary behavior of a given model is described in terms of a generalized slow-roll attractor solution in phase space, we have therefore looked at possible on-shell transformations of a given non-canonical model on its inflationary attractor into an equivalent canonical slow-roll inflation model. We have constructed such on-shell transformations in general, so that given a non-canonical lagrangian which supports inflation, the potential required to reproduce the inflationary trajectory $X_{inf}(\phi)$ in a canonical theory can be found. 
Furthermore, we checked for the matching of the 2-point function observables $\Delta_s^2$, $n_s$, and $r$. We find a full on-shell match for all 2-point function quantities precisely for the case of DBI inflation while failing for the DBI-inspired generalizations. This can be shown analytically and numerically. This may point to a special status of DBI inflation as a member of the non-canonical class in that it can be related to a canonical model of inflation with matching 2-point function observables..

Lastly, in the light of the much-awaited Planck data on nongaussianity, we would like to point out that given the data we have, there remains a large degree of degeneracy between inflationary models, which we have to bear in mind when interpreting that data. Since it is often claimed that canonical and noncanonical theories can be distinguished using the data, we feel this added degeneracy serves as a warning that this may not be the case, particularly if large NG is not observed. Unless data on nongaussianities improves drastically and reveals a non-negligible single of equilateral nongaussianity, or the consistency relation between $r$ and $n_T$ can be accurately measured, one may never be able to distinguish between non-canonical inflation and slow-roll inflation in some canonical theory. In fact even with the observation of NG, this differentiation may not be possible: Note that appreciable non-Gaussianity can arise in single scalar field theories of inflation with a canonical kinetic term, from features 
in the potential \cite{Gwyn:2012pb}, or from coupling of the inflaton to gauge quanta \cite{Barnaby:2010vf, Barnaby:2011qe}. It is possible that by adding additional couplings or features to the potential of the canonical theory one could match observables at the 3-point function level as well. We have not addressed the question of matching 3-point observables such as non-Gaussianity here, and leave investigation of this question for future work. Also the deeper reason for the agreement of DBI with its canonical transform at the level of the 2-point function is yet to be understood on a more fundamental level. This degeneracy thus opens many questions for future study.


\acknowledgments We thank Jan Louis, Raquel Ribeiro and especially Bret Underwood for valuable, and enlightening discussions. This work was supported by the Impuls und Vernetzungsfond of the Helmholtz Association of German Research Centers under grant HZ-NG-603, the German Science Foundation (DFG) within the Collaborative Research Center 676 ``Particles, Strings and the Early Universe" and the Research Training Group 1670. R.G. is grateful for support by the European Research Council via the Starting Grant numbered 256994. R.G. was also supported during the initial stages of this work by an SFB fellowship within the Collaborative Research Center 676 ``Particles, Strings and the Early Universe'' and would like to thank the theory groups at DESY and the University of Hamburg for their hospitality at this time.

\appendix
\section{Accessing the non-canonical regime in DBI} \label{sec:AppA}
Here we comment on constraints on the allowed phase space of brane inflation models governed by the DBI action eq.~(\ref{DBIinflp}). It was shown in \cite{Franche:2010yj} that the presence of non-canonical kinetic terms can ameliorate the initial conditions fine tuning problem, but this effect is only present when the non-canonical terms are relevant in the allowed phase space.  The non-canonical kinetic terms are relevant when $A \geq 1$ or  $|\Pi| > \Lambda^2$\cite{Franche:2010yj}, where $\Pi$ is the canonical momentum given by 
\begin{equation}
\Pi = - \sqrt{2X} \frac{\partial p(X, \phi)}{\partial X}. 
\end{equation}
Clearly, however, the available phase space is bounded. For an effective field theory description to be valid, we require $H < \Lambda$, which implies $|\Pi| < M_p \Lambda$ for a canonical kinetic term and $|\Pi| < M_p^2$ for a non-canonical kinetic term. This leaves a potentially large range of momenta $\Lambda^2 < \Pi \leq M_p^2$ in which the non-canonical kinetic terms are relevant and the EFT description remains valid. In addition, there can be bounds on the range of the inflaton field $\phi$ which restrict the allowed phase space.

The DBI action \cite{Silverstein:2003hf} describes inflation in the D3/$\overline{D3}$ inflationary setup \cite{Kachru:2003sx},  in which a mobile D3 brane moves in a warped throat in the internal (compactified) space. The inflaton field is related to the brane separation, and governed by eq.~(\ref{DBIinflp}) where $f(\phi)$ is the warp factor of the throat in the internal space.  Inflation proceeds as the D3 brane moves towards an $\overline{D3}$-brane at the bottom of the throat, ending when the branes are so close together that the strings stretching between them become tachyonic and the branes annihilate. This scenario has been widely studied - see, for instance,  \cite{Baumann:2007ah,  Baumann:2008kq, Baumann:2009qx, Bena:2010ze, Agarwal:2011wm}. In this setup, there are both upper and lower bounds on the inflaton field $\phi$:
\begin{equation}
 \mu \ll \phi \ll M_p\,.
\end{equation}
A lower limit can be understood physically from the requirement that the branes must be initially separated by at least a string length so that inflation does not end immediately. The upper limit reflects the fact that the inflaton range cannot be larger than the size of the compactified space. 

Critically, the upper bound on the field range is related to the warp factor.  Along with the lower bound on the field, this implies an upper bound on the warp factor $f(\phi) = \frac{\lambda}{\phi^4}$ and therefore the `non-canonicalness' parameter \cite{Franche:2009gk}
\begin{equation}
A_{DBI} \equiv \frac{V'(\phi)f^{1/2}(\phi)}{3 H(\phi)}.
\end{equation}
This can be understood as follows \cite{Baumann:2006cd, Peiris:2007gz, Bird:2009pq}: the 4-dimensional Planck mass scales with the warped volume of the compact space, and is therefore an upper bound on the volume of the warped throat:
\begin{eqnarray}
\frac{\mathrm{Vol(X_5)}}{\pi} \int_{\phi_{end}}^{\phi_{UV}} d \phi \phi^5 f(\phi) & < &  M_p^2.
\end{eqnarray}
Using $\lambda = \frac{1}{2} \frac{N \pi}{\mathrm{Vol(X_5)}}$ from the Klebanov-Strassler throat solution~\cite{Klebanov:2000hb}, where $N$  is the amount of 5-form flux associated with the warping, this gives rise to the bounds
\begin{eqnarray}
\pi^2 \lambda \phi_{UV}^2/2 &<&M_p^2, \\
\label{fieldrangebound} (\Delta \phi/ M_{pl})^2&< &\frac{4}{N},\\
\label{fbound}\sqrt{f} & <& \frac{\sqrt{2}M_P}{\pi \phi_{UV} \phi^2},
\end{eqnarray}
where we took $\mathrm{Vol(X_5)} \sim \pi^3$ as in \cite{Peiris:2007gz}. 
We find the field range bound eq.~(\ref{fieldrangebound}), or equivalently a bound on the warp factor eq.~(\ref{fbound}).  Since $\phi$ also has a lower limit, there is immediately an upper bound on $f$. The lower limit is found to be $\phi^4 \approx D$ \cite{Bird:2009pq}, where
\begin{equation}
 D \sim  \frac{T_3}{ h_{IR}}  \sim \frac{ m_s^4} {g_s h_{IR}}\,.
\end{equation}
This gives us a bound on $A_{DBI}$:
\begin{eqnarray}
A_{DBI} & < & \frac{\sqrt{2} V'}{\sqrt{3} \pi \phi_{UV} \phi^2 \sqrt{V}},
\end{eqnarray}
where we have set $M_p = 1$.

Using the potential \cite{Bird:2009pq}
\begin{equation}
V(\phi) = Ds(1 - \frac{CD}{\phi^4} +  \alpha \phi + \frac{\beta \phi^2}{3} - a_{\Delta} \phi^\Delta),
\end{equation}
where $s$ is some number of order 1, and $C = \frac{3}{16 \pi^2 s} \ll 1$,  $a_\Delta = 0$ for a quadratic potential and $\Delta = \frac{3}{2}$ for an inflection point potential, the bound becomes~\cite{Bird:2009pq}
\begin{eqnarray}
\begin{aligned}
A &<& \sqrt{\frac{2}{3}}\frac{\sqrt{Ds}}{\pi \phi_{UV} \phi^2}\frac{(\frac{2 \beta \phi}{3} + \frac{4CD}{\phi^5})}{\sqrt{1 + \alpha \phi + \frac{\beta \phi^2}{3} - a_{\Delta} \phi^\Delta - \frac{CD}{\phi^4}}}\,,\\
A & <& \sqrt{\frac{2}{3}}\frac{\sqrt{Ds}}{\pi \phi_{UV} \phi^2}\frac{(\frac{2 \beta \phi}{3} + \frac{4CD}{\phi^5})}{\sqrt{1 + \alpha \phi+ \frac{\beta \phi^2}{3} - a_{\Delta} \phi^\Delta - \frac{CD}{\phi^4}}}\,.
\end{aligned}
\end{eqnarray}

It is only if $A$ can be larger than 1 within the allowed field range of $\phi$ that the non-canonical regime will be accessed. We can test whether this is the case by considering three cases. Note that $\frac{D}{\phi^4}$ has upper bound $1$. 
\begin{itemize}
\item {\bf Case One: $\frac{D}{\phi^4} \ll 1$}. 
In this case we are firmly in the canonical regime where $A \ll 1$: 
\begin{eqnarray}
\begin{aligned}
 A & \lessim \frac{\sqrt{\frac{D}{\phi^4} s} ( 2 \phi)}{\phi_{UV} \pi} \,,\\
A & \lessim \sqrt{\frac{D}{\phi^4}} \frac{\phi}{\phi_{UV}}\,,\\
A & \ll 1.
\end{aligned}
\end{eqnarray}
\item {\bf Case Two: $\frac{D}{\phi^4} \sim 1$}:  In this case A can be very large because the denominator blows up. However, $\phi \approx \phi_{end}$ so by the time this happens inflation is already over. 
\item {\bf Case Three: $\frac{D}{\phi^4}$ is intermediate.}
Let us be more careful about the approach to large $A$, taking  $\frac{D}{\phi^4}$ to be some intermediate value $\leq 1$. In this case 
\begin{equation}
A \lessim \sqrt{\frac{2s}{3}} \frac{4C}{\pi} \frac{D^{3/2}}{\phi^7 \phi_{UV}} + \left (\frac{2}{3} \right)^{3/2} \sqrt{\frac{Ds}{\phi^4} } \frac{\beta \phi}{\pi \phi_{UV}}.
\end{equation}
 Note that the denominator we have taken to be 1 will generally be less than one for this intermediate case, thus weakening the bound. Each factor in the second term is smaller than 1, so it will only be possible to get a contribution to $A$ of order 1 or greater from the first term. Then
\begin{eqnarray}
\begin{aligned}
A & \sim & \sqrt{\frac{2s}{3}} \frac{4C}{\pi} \frac{D^{3/2}}{\phi^7 \phi_{UV}}\,,\\
A & \sim & \sqrt{\frac{2s}{3}} \frac{4C}{\pi} \frac{\phi_{IR}^6}{\phi^7 \phi_{UV}}\,.
\end{aligned}
\end{eqnarray}
Taking $C = \frac{3}{16 \pi^2 s}$, $s \approx 1$,  $\phi = a \phi_{IR}$, where $a$ is some positive number greater than 1, this gives
\begin{eqnarray}
\begin{aligned}
A & \sim 0.02 \frac{\phi_{IR}^6 }{\phi^7 \phi_{UV}}\,,\\
A & \sim 0.02 \frac{1}{a^7 \phi_{IR} \phi_{UV}}\,.
\end{aligned}
\end{eqnarray}

Then we see that there is a relation between how small $\phi_{IR}$ can be and how far from the end of inflation (parametrized by $a$) we can access the non-canonical regime. For a fixed $A \geq 1$, decreasing $\phi_{IR}$ implies a larger $a$ from which we access the non-canonical regime, but the fraction of the field range in the NC regime is reduced. For $\phi_{UV} = 0.1$ and  $\phi_{IR} = 0.01$, $a \leq 1.53$, giving a very small range of $\phi$ values for which $A \geq 1$:
\begin{equation}
\frac{\phi - \phi_{IR}}{\phi_{UV} - \phi_{IR} } \,\, = \frac {(a-1) \phi_{IR}}{\phi_{UV} - \phi_{IR}} \,\, \leq \,\,  0.06.
\end{equation}
 Increasing $\phi_{IR}$ will increase this fraction, but of course one must have $\phi_{IR} \ll \phi_{UV} \ll 1$ for consistency. For $\phi_{UV} = 0.1$ and  $\phi_{IR} = 0.03$, $a \leq 1.2$, so that $A \geq 1$ for 13\% of the field range.  Thus it is possible for A to exceed 1, although not parametrically, in a realistic D3/$\overline{D3}$ setup. It thus seems that NCI inflation can occur in this setup only for some short range of $\phi$ values, consistent with the conclusions of \cite{Baumann:2006cd,Bean:2007hc, Bird:2009pq}. 
\end{itemize}

\section{Off-shell canonical transformations} \label{sec:AppB}
In this article we have used the attractor equation to make an on-shell transformation from a theory with non-canonical kinetic term to a theory with canonical kinetic term, such that the inflationary trajectory is the same. Off-shell transformations, in which the equation of motion is not used, can also be used to transform between such theories, but have a more limited scope.

Canonical transformations were used in \cite{Bean:2008ga} to transform from theories with canonical kinetic terms to theories with non-canonical kinetic terms, in  $0+1$ dimensions. We shall see that this is possible when, in the case that the Lagrangian is separable, the form of the kinetic and potential terms are exchanged by the canonical transformations. 

A canonical transformation is defined by a generating function $F(\phi,\tphi)$ via
\begin{equation}
 p = \frac{\del F}{\del \phi}\,,\qquad \qquad \tilde p = - \frac{\del F}{\del \tphi}\,.\label{ptildep_def}
\end{equation}

As an example, let $L  =  \frac{1}{2} \dot \phi^2 - V(\phi)$, $V(\phi) = k \phi^{4/3}$ and $F(\phi, \tilde \phi, t) = \phi f(\tilde \phi)$. Then the transformations can be written
\begin{eqnarray}
\begin{aligned}
p & =  f(\tilde \phi)\,,\\
\phi & =  - \frac{\tilde p}{f'(\tilde \phi)}.
\end{aligned}
\end{eqnarray}

We have the energy density $H = \tilde H$ and can find $\dot {\tilde \phi} = \frac{\partial \tilde H}{\partial \tilde p}$:
\begin{eqnarray}
\begin{aligned}
\tilde H & = \frac{1}{2} f(\tilde \phi)^2 + k \left (- \frac{\tilde p}{f'(\tilde \phi)} \right ) ^{4/3}\,, \\
\dot {\tilde \phi} & = - \frac{4k}{3 f'(\tilde \phi)} \left ( - \frac{\tilde p}{f'(\tilde \phi)}\right )^{1/3} \,.
\end{aligned}
\end{eqnarray}
Invert this to get $\tilde p = \frac{3^3}{4^3} \frac{(f')^4}{k^3} \dot {\tilde \phi}^3$. Then the transformed Lagrangian is found to be:

\begin{eqnarray}
\begin{aligned}
\tilde L & =  \tilde p \dot {\tilde \phi} - \tilde H\,,\\
\tilde L& =  \frac{3^3}{4^4} \frac{(f')^4}{k^3} \dot{\tilde \phi}^4 - \frac{1}{2} f^2\,,
\end{aligned}
\end{eqnarray}
which has a non-canonical kinetic term $X^2 \sim \dot \tilde \phi^4$. We can now ask what the general conditions are for it to be possible to obtain an action with canonical kinetic term upon performing a canonical transformation. 

\subsubsection*{General conditions for the existence of a dual canonical theory}

We start with a separable Hamiltonian
\begin{equation}
 H(p,\phi) = K(p) + V(\phi)\,,
\end{equation}
with kinetic term $K(p)$ and potential $V(\phi)$.
To obtain the transformed Hamiltonian one has to invert the second relation in eq.~\eqref{ptildep_def} to find the dependencies
\begin{equation}
 p = p(\tp,\tphi)\,,\qquad \qquad \phi = \phi(\tp,\tphi)\,.
\label{pphifunc}
\end{equation}
The transformed Hamiltonian is then given as
\begin{equation}
 \tilde H(\tilde p,\tphi) = K(p(\tp,\tphi))+V(\phi(\tp,\tphi))\,.
\label{Htildegen}
\end{equation}
This is generally not a separable Hamiltonian, i.e. of the form $\tilde K(\tp) + \tilde V(\tphi)$, let alone in canonical form.

By Taylor expanding the transformed Hamiltonian around $\tp=0$:
\begin{equation}
 \tilde H(\tilde p,\tphi) = \tilde H(0,\tphi) + \frac{\del^2 \tilde H}{\del \tp^2}_{|\tp=0} \frac{\tp^2}{2} + \sum_{i=1,i\neq 2}^\infty \epsilon_i \tp^i\,,\quad \text{with}\quad \epsilon_i = \frac{1}{i !}\frac{\del^i \tilde H}{\del \tp^i}_{|\tp=0}\,,
\end{equation}
we see that the transformed theory is approximately canonical with potential $\tilde V(\tphi) = \tilde H(0,\tphi)$ iff the generating function can be chosen such that
\begin{equation}
 \frac{\del^2 \tilde H}{\del \tp^2}_{|\tp=0} = 1 \qquad \text{and} \qquad |\epsilon_i| \ll 1\,.
\end{equation}

\subsubsection*{Simplifying approach $K \leftrightarrow V$} 

The simplest way to obtain a separable dual theory is to demand that the transformations eq.~\eqref{pphifunc} exchange the role of the kinetic term $K(p)$ and the potential $V(\phi)$. This happens if $p(\tp,\tphi)$ is only a function of $\tphi$ and $\phi(\tp,\tphi)$ is only a function of $\tp$.

This requirement determines the form of the generating function $F(\phi,\tphi)$: First, $p=p(\tphi)=\frac{\del F(\phi,\tphi)}{\del \phi}$ determines $F(\phi,\tphi)$ to be linear in $\phi$, i.e. $F(\phi,\tphi)=a(\tphi)+b(\tphi) \phi$. Second, $\tp = - \frac{\del F}{\del \tphi} = -a'(\tphi)-b'(\tphi) \phi$ will only give a relation $\phi = \phi(\tp)$ independent of $\tphi$ if $a'(\tphi)$ and $b'(\tphi)$ do not depend on $\tphi$. Hence, the most general $F(\phi,\tphi)$ that exchanges the role of $K$ and $V$ can be parametrized as
\begin{equation}
 F(\phi,\tphi)=(k\phi+g)\tphi\,,\qquad \text{with} \qquad k,g\in \mathbb{R}\,, k\neq0\,.
\end{equation}
The transformation is then linear and given by
\begin{equation}
 p = k\tphi \,,\qquad \qquad \phi = \frac{-g-\tp}{k}\,.
\label{pphiKV}
\end{equation}
We will apply this type of generating function in the following sections ~\ref{DBIm2phi2_sec} and~\ref{DBIinflection_sec} 
to obtain a canonical theory that is dual to a non-canonical theory.

\subsection{DBI + quadratic potential} \label{DBIm2phi2_sec}

The Hamiltonian is given by
\begin{equation}
 H(p,\phi) = \frac{1}{f} \left(\sqrt{1+f\,p^2} -1 \right) + V(\phi)\,,\qquad \text{with} \qquad V(\phi) = \frac{1}{2} m^2 \phi^2\,. \label{HDBIm2phi2}
\end{equation}
For the generating function $F=m\phi \tphi$, this becomes the canonical theory
\begin{equation}
 \tilde H(\tp,\tphi) = \frac{1}{2}\tp^2 + \tilde V(\tphi)\,,\qquad \text{with} \qquad \tilde V(\tphi) = \frac{1}{f} \left(\sqrt{1+f m^2\,\tphi^2} -1 \right)\,. \label{HcanDBIm2phi2}
\end{equation}

The parameter that indicates if the theory is in the canonical or non-canonical regime is
\begin{equation}
 A = \frac{V'(\phi) f^{1/2}}{3 H} \simeq \frac{V'(\phi) f^{1/2}}{\sqrt{3 V}}\,,\label{Adef}
\end{equation}
where in the last step we have used $H^2 \simeq V/3$, i.e. the energy is dominated by the potential energy. If $A\gg 1$ the theory is in the non-canonical regime, while for $A \ll 1$ it is in the canonical regime.

For the non-canonical DBI theory given in eq.~\eqref{HDBIm2phi2}, the A-parameter is given as
\begin{equation}
 A_{DBI} = \sqrt{\frac{2 f}{3}} \, m\,.
\end{equation}
Hence, to be in the non-canonical regime we have to demand $\sqrt{f}\, m\gg 1$. In this limit, we can approximate the potential $\tilde V(\tphi)$ of the canonical theory eq.~\eqref{HcanDBIm2phi2} by
\begin{equation}
 \tilde V(\tphi) = \frac{m \tphi}{\sqrt{f}}\,.
\end{equation}

\subsection{DBI + ``inflection point potential''} \label{DBIinflection_sec}

We now want to look at the theory
\begin{equation}
H(p,\phi) = \frac{1}{f} \left(\sqrt{1+f\,p^2} -1 \right) + V(\phi)\,,\quad \text{with} \quad V(\phi) = V_0 + \lambda (\phi-\phi_0) + \beta (\phi - \phi_0)^3\,. \label{HDBIinfl}
\end{equation}
This potential is suitable for small field inflation since for $\lambda, \beta \ll V_0$ the slow-roll parameters $\epsilon$ and $\eta$ are small at $\phi = \phi_0$, without the necessity of $\phi$ having to travel a trans-Planckian distance as for instance in chaotic inflation.

To obtain a potential where the inflaton is rolling down towards a local minimum we have to choose $\lambda,\beta>0$. Eq.~\eqref{HDBIinfl} then only describes the dynamics near the inflection point $\phi=\phi_0$.

\begin{figure}[h!]
\centering
\includegraphics[width= 0.65\linewidth]{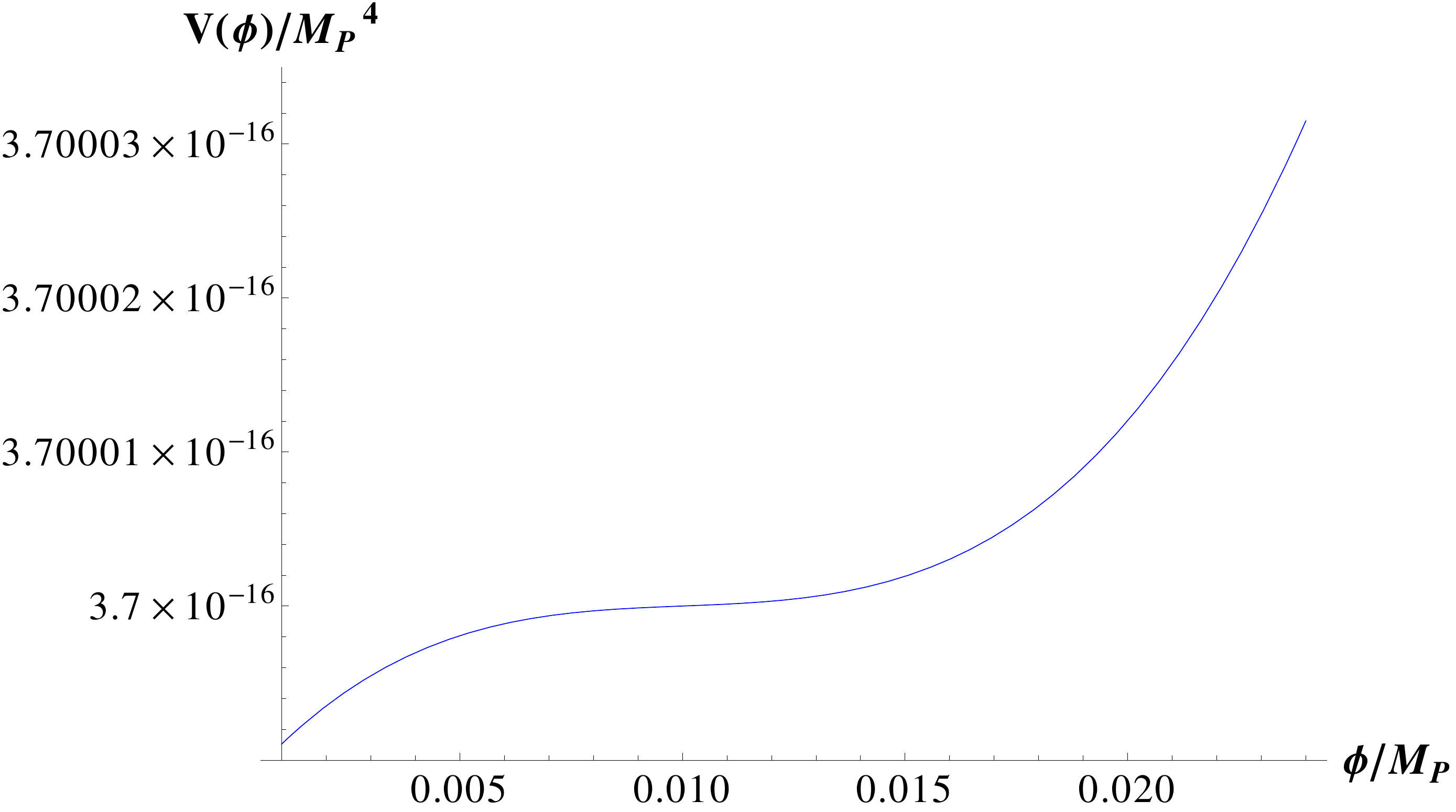}
\caption{Inflection point potential, eq.~\eqref{HDBIinfl} for the parameters $V_0 = 3.7 \cdot 10^{-16}$, $\lambda = 1.13\cdot 10^{-20}$, $\beta = 1.09\cdot10^{-15}$ and $\phi_0 = 0.01$.}
\label{inflectionpot_fig}
\end{figure}
The $A$-parameter~\eqref{Adef} for the theory described by eq.~\eqref{HDBIinfl} generically demands $f\gg1$ for the theory to be in the non-canonical regime.

Using the generating function $F(\phi,\tphi)=(k\phi+g)\tphi$ we can transform the Hamiltonian eq.~\eqref{HDBIinfl} into the approximately canonical theory
\begin{equation}
 \tilde H(\tp,\tphi) = \frac{1}{2}\tp^2 + \epsilon_1 \tp + \epsilon_3 \tp^3 + \tilde V(\tphi)\,, \label{HcanDBIinfl}
\end{equation}
with
\begin{equation}
 \tilde V(\tphi) = \frac{1}{f} \left(\sqrt{1+f k^2 \tphi ^2} -1 \right)+V_0-\frac{1}{108 \epsilon _3^2}+\frac{\epsilon _1}{6 \epsilon _3}\,,
\end{equation}
where we have used 
\begin{equation}
g = -k \phi_0 +\frac{1}{6 \epsilon_3}\,,\quad \beta = -k^3 \epsilon_3\,,\quad \lambda = -k \epsilon_1+\frac{k}{12 \epsilon_3} \,. \label{gbetalambdarep}
\end{equation}
The first equation in~\eqref{gbetalambdarep} follows from the canonical normalization of the $\tp^2$ term in eq.~\eqref{HcanDBIinfl} while the other two equations are reparametrizations of the potential parameters $\lambda$ and $\beta$ such that the coefficients of $\tp$ and $\tp^3$ are small in eq.~\eqref{HcanDBIinfl}, i.e. $\epsilon_1, \epsilon_3 \ll 1$. At this point, $k$, $V_0$ and $\phi_0$ remain free unfixed parameters.

We see that for $\epsilon_1, \epsilon_3 \ll 1$ it is impossible for $\lambda$ and $\beta$ to have the same sign since $\beta \sim -k^3 \epsilon_3$ and $\lambda \sim k/ 12 \epsilon_3$. Hence, the original idea of the inflection point inflaton potential visualized in Figure~\ref{inflectionpot_fig} does not have a dual canonical theory that is related via a generating function $F(\phi,\tphi)=(k\phi+g)\tphi$.

\begin{figure}[h!]
\centering
\includegraphics[width= 0.49\linewidth]{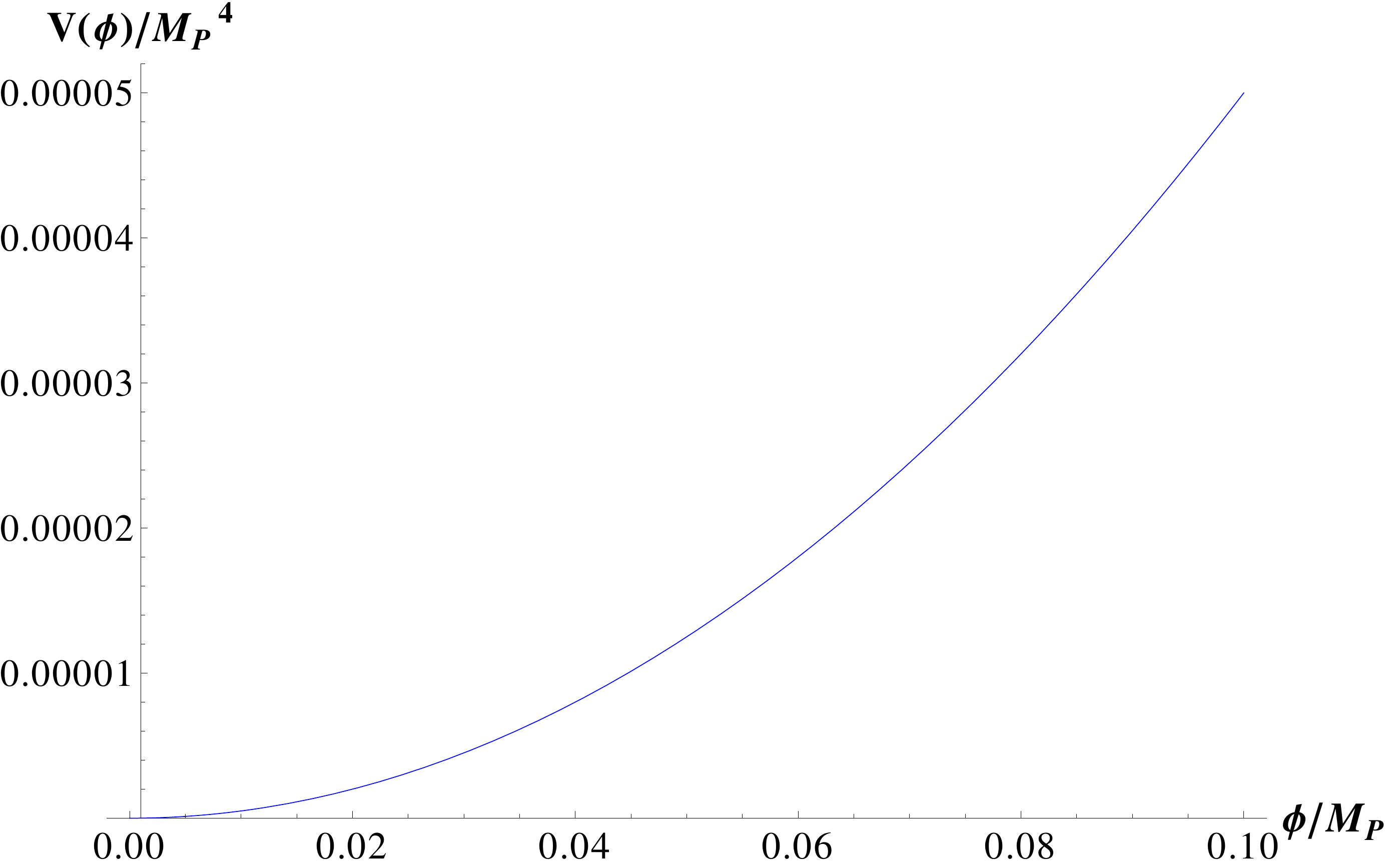}
\includegraphics[width= 0.49\linewidth]{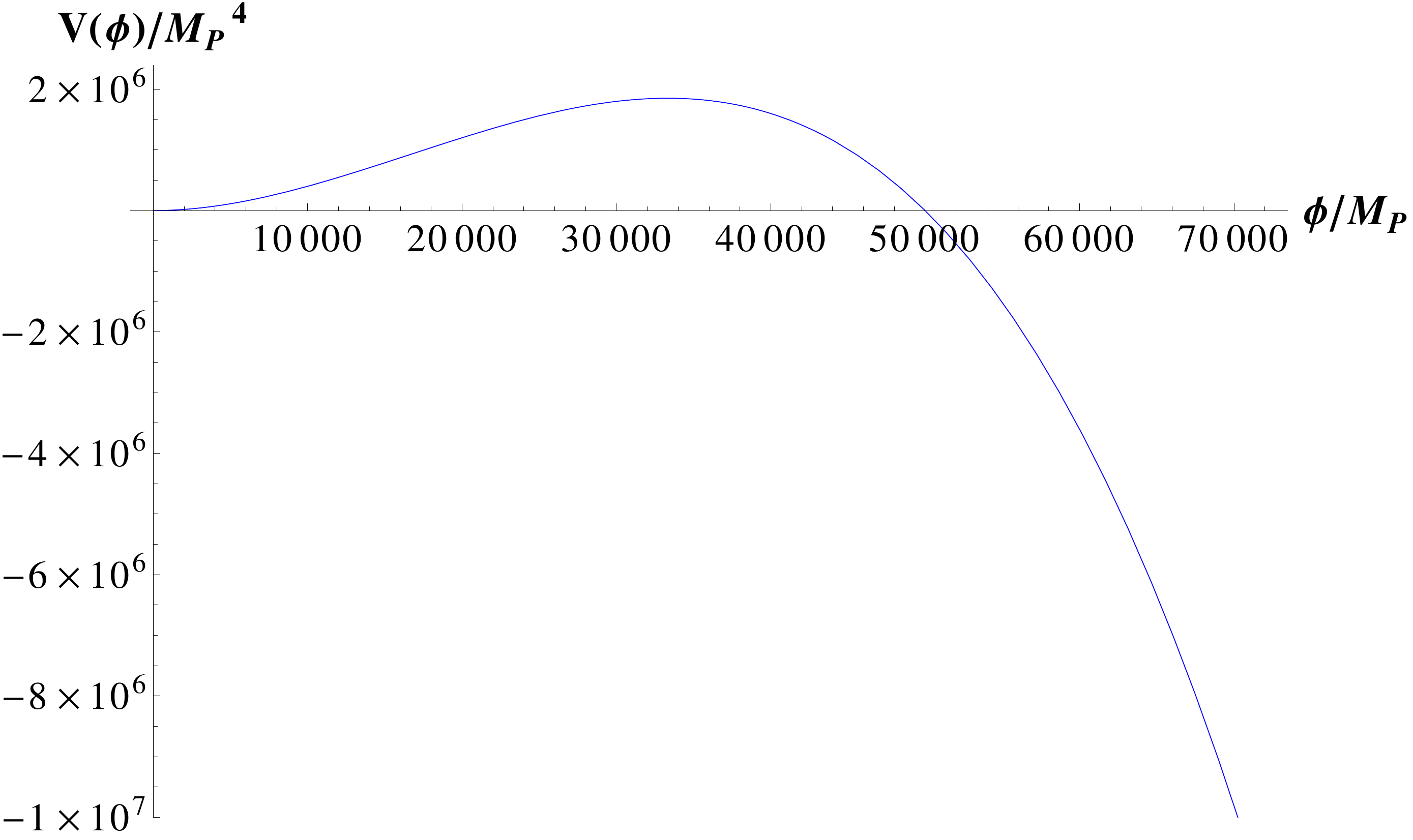}
\caption{Modified inflection point potential, eq.~\eqref{modinflpot} for the parameters $k=0.1$, $\epsilon_1=\epsilon_3=10^{-4}$.}
\label{inflectionpotmod_fig}
\end{figure}

However, we can still go ahead and try to construct a sensible inflationary theory on both the canonical and non-canonical side respecting the just discussed constraints on $\lambda$ and $\beta$. In the non-canonical theory $H(p,\phi)$ we choose $\lambda>0$ and hence $\beta<0$ in order for the potential to have $V'(\phi)<0$, i.e. the inflaton is rolling down towards smaller field values. Furthermore, we demand the local minimum of the potential to be at $\phi=0$ and $V(\phi)=0$ which fixes $\phi_0$ and $V_0$ to be
\begin{equation}
 \phi_0 = \frac{\sqrt{1-12 \epsilon _1 \epsilon _3}}{6 k \epsilon _3}\,,\qquad V_0 = \frac{\left(1-12 \epsilon _1 \epsilon_3\right)^{3/2}}{108 \epsilon _3^2}\,.\label{phi0V0rep}
\end{equation}
Inserting eqs~\eqref{phi0V0rep} into $V(\phi)$ the potential simplifies to
\begin{equation}
 V(\phi)= \frac{1}{2} k^2 \phi^2 \left(\sqrt{1-12 \epsilon _1 \epsilon _3} -2 k \epsilon _3 \,\phi\right) \simeq \frac{1}{2} k^2 \phi^2\,,\label{modinflpot}
\end{equation}
i.e. an approximately quadratic potential with mass parameter $k$, see also Figure~\ref{inflectionpotmod_fig}.

In other words, our demand that there exists an approximately canonical dual theory implies that the potential of the non-canonical theory we started from is approximately quadratic and we are back to our discussion in section~\ref{DBIm2phi2_sec}.

To complete the analogy with section~\ref{DBIm2phi2_sec}, we look at the potential $\tilde V(\tphi)$ of the approximately canonical theory, inserting the expression~\eqref{phi0V0rep} for $V_0$:
\begin{align}
 \tilde V(\tphi) &= \frac{1}{f} \left(\sqrt{1+f k^2 \tphi ^2} -1 \right) -\frac{1}{108 \epsilon _3^2}+\frac{\epsilon _1}{6 \epsilon _3}+\frac{\left(1-12 \epsilon _1 \epsilon _3\right){}^{3/2}}{108 \epsilon _3^2}\notag\\
 &= \frac{1}{f} \left(\sqrt{1+f k^2 \tphi ^2} -1 \right) + \order(\epsilon_1^2,\epsilon_3)\,.
\end{align}
In the limit $\epsilon_1,\epsilon_3\to 0$, this is identical with the potential found in eq.~\eqref{HcanDBIm2phi2}.
 
Our attempts to transform a theory with non-canonical kinetic term and inflection point potential to an approximately canonical theory clearly show the limitations of the $K\leftrightarrow V$ ansatz. From this example, we conclude that the $K\leftrightarrow V$ ansatz is only useful in the case of a potential with a dominant quadratic term in the non-canonical theory and in this case there always exists a dual canonical theory. For non-quadratic potentials one has to go beyond the $K\leftrightarrow V$ ansatz.

\newpage
\bibliographystyle{JHEP.bst}
\bibliography{CNCJCAP}
\end{document}